%% file: coincidence_2026_arxiv_revise.tex
\documentclass[pra,floatfix,twocolumn,showpacs,amsmath,amssymb]{revtex4}
\usepackage{graphicx}
\usepackage{txfonts}
\usepackage{bm}
\usepackage{color}
\usepackage{mathrsfs}

\begin{document}

\title{Narrow-Sense Type-III Dirac Cones and Additional Flat Lines on a Honeycomb Lattice with Anisotropic Next-Nearest-Neighbor Hoppings}

\author
{Keita Kishigi}
\affiliation{Faculty of Education, Kumamoto University, Kurokami 2-40-1, 
Kumamoto, 860-8555, Japan}



\author{Yasumasa Hasegawa}
\affiliation{Department of Material Science, 
Graduate School of Science, 
University of Hyogo, Hyogo, 678-1297, Japan}

\begin{abstract}

Critically tilted Dirac cones have attracted considerable attention because of their unconventional electronic properties in two-dimensional massless Dirac-fermion systems. Among them, the narrow-sense type-III Dirac cone is characterized by a flat dispersion along the direction connecting the two Dirac points.

Using a tight-binding model on a single-orbital honeycomb lattice, we demonstrate that narrow-sense type-III Dirac cones can be realized by tuning  anisotropic next-nearest-neighbor hoppings. Type-I and type-II Dirac cones emerge on either side of the critical point, enabling a systematic investigation of the electronic properties across the type-I, narrow-sense type-III, and type-II regimes.

We further find that the present model exhibits additional flat lines in momentum space whose energy coincides with the Dirac-point energy at the narrow-sense type-III critical point. These flat lines lead to a pronounced enhancement of the density of states above the Fermi energy and, consequently, the electronic specific heat is substantially enhanced compared with that expected for an ideal narrow-sense type-III Dirac cone. Our results show that the honeycomb lattice with anisotropic next-nearest-neighbor hoppings provides a simple platform for exploring unconventional thermodynamic properties arising from the energy coincidence of narrow-sense type-III Dirac cones and additional flat lines at the Fermi energy. 

\end{abstract}



\date{\today}


\maketitle

\section{Introduction}

\begin{figure}[bt]
\begin{flushleft} \hspace{0.5cm}(a) \end{flushleft}\vspace{-0.2cm}\hspace{0.2cm}
\includegraphics[width=0.41\textwidth]{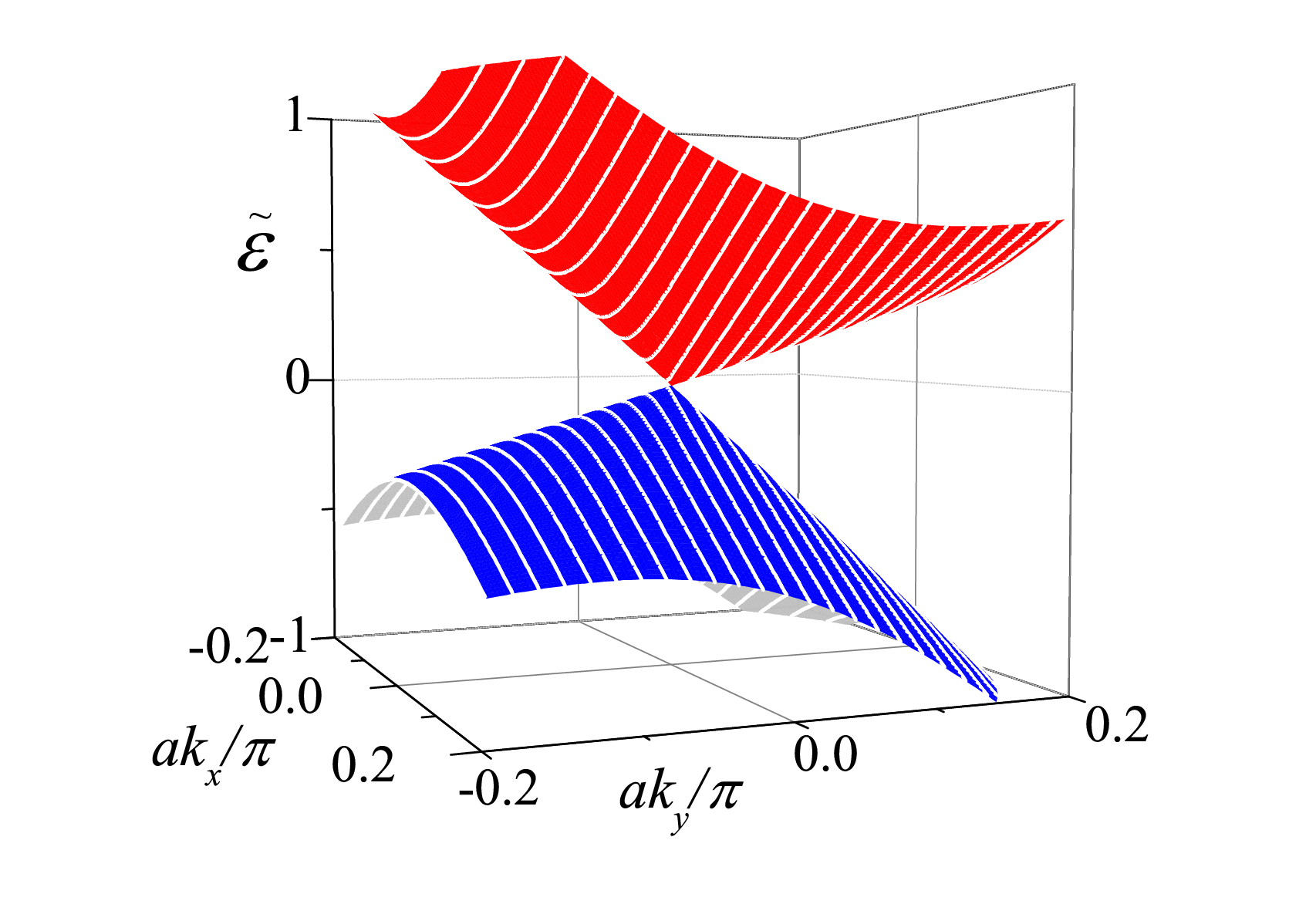}\vspace{-0.2cm}
\begin{flushleft} \hspace{0.5cm}(b) \end{flushleft}\vspace{-0.2cm}\hspace{0.2cm}
\includegraphics[width=0.41\textwidth]{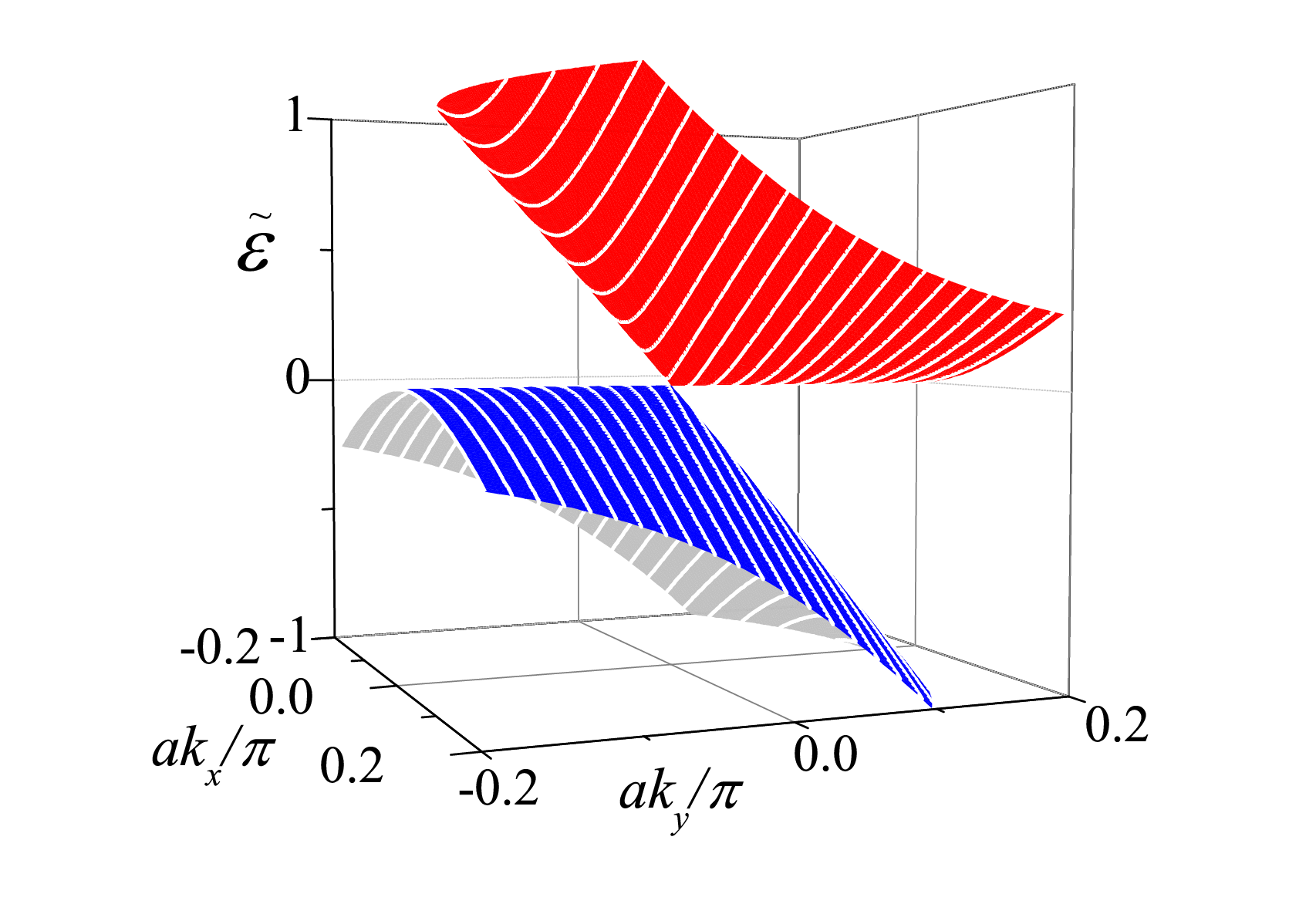}\vspace{-0.2cm}
\begin{flushleft} \hspace{0.5cm}(c) \end{flushleft}\vspace{-0.2cm}\hspace{0.2cm}
\includegraphics[width=0.41\textwidth]{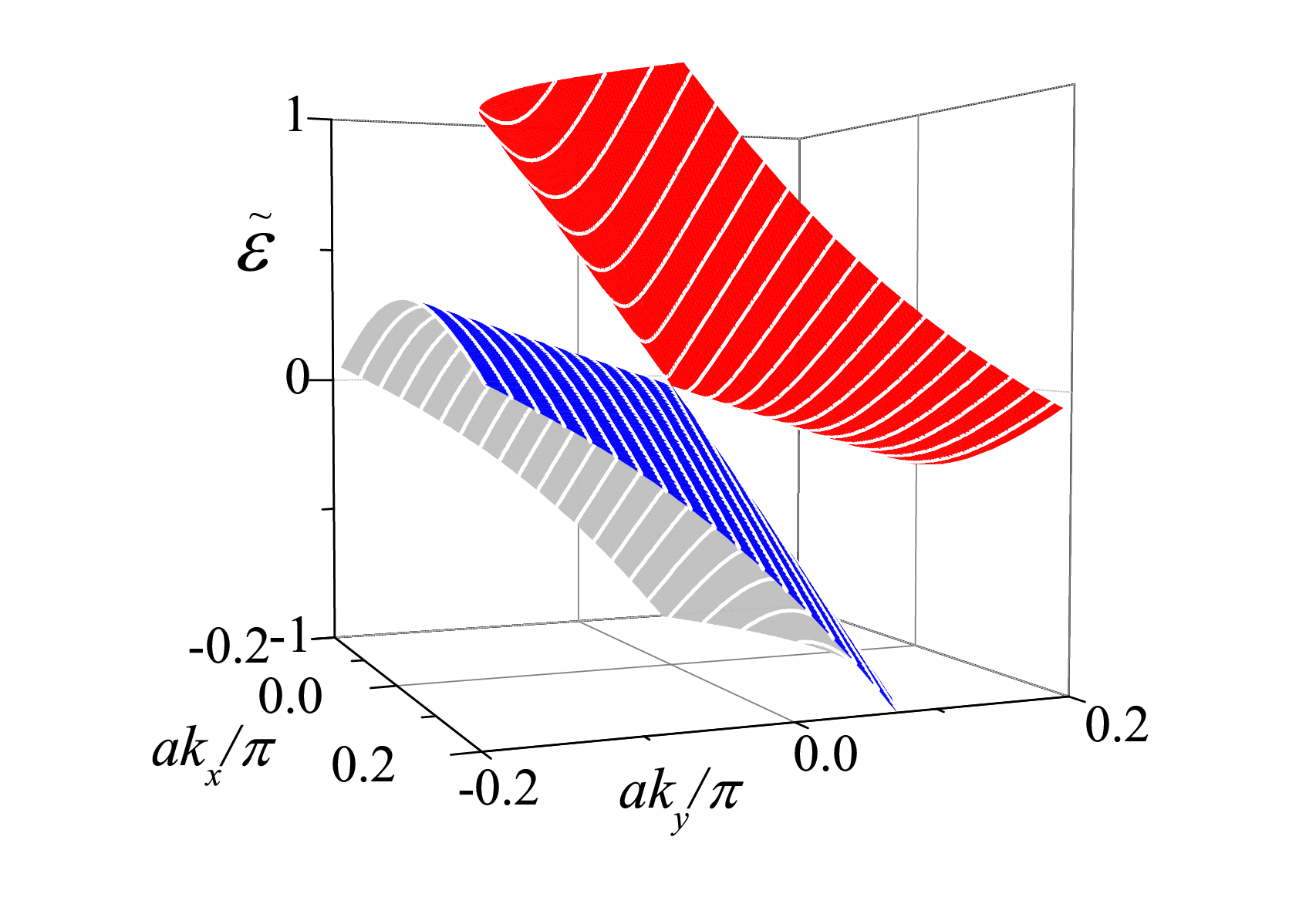}\vspace{-0.2cm}
\caption{(Color online) 
The energy bands of 2D type-I (a), narrow-sense type-III (b), and type-II (c) Dirac, which are described by Eq. (\ref{model}), where ${\tilde \varepsilon} = a(\varepsilon({\pm}, k_x, k_y)-{\tilde \varepsilon} _{\rm F})/(\pi \hbar v_{\rm F})$ and $a$ represents the lattice constant. 
The tilting is classified as type-I for $|\alpha/(\hbar v_{\rm F})| < 1$, narrow-sense type-III for $|\alpha/(\hbar v_{\rm F})| = 1$, and type-II for $|\alpha/(\hbar v_{\rm F})| > 1$. We take $\alpha/(\hbar v_{\rm F})=-0.5$ for (a), $\alpha/(\hbar v_{\rm F})=-1$ for (b), and 
$\alpha/(\hbar v_{\rm F})=-1.5$ for (c).
}
\label{fig1}
\end{figure}
%
%

Two-dimensional (2D) massless Dirac fermions have been realized in graphene \cite{novo2005}, $\alpha$-(BEDT-TTF)$_2$I$_3$ \cite{Kajita2014,Hirata2011,Osada2008,Konoike2012}, and $\alpha$-(BETS)$_2$I$_3$ \cite{Tajima2021,kitou2021}, where the electronic band structures exhibit Dirac points with linear dispersions. When two Dirac points merge, a topological phase transition from a semimetallic state to an insulating state can occur, and the merging point is referred to as a semi-Dirac point \cite{Bane}.
Such a transition has been theoretically predicted in various systems \cite{Hasegawa2006,Dietl2008,Bane,Suzumura2013} and experimentally realized in artificial systems \cite{Tarruell2012,Bellec,Mil,real}, as well as in real materials such as ZrSiS \cite{Shao}. Near the merging of two Dirac points, the density of states (DOS) changes from Dirac-like to semi-Dirac-like behavior, leading to a crossover in the temperature ($T$)-dependence of the electronic specific heat ($C$)\cite{KTH2023}.

The concept of a tilted Dirac cone has been extensively studied in 2D massless Dirac systems, particularly in $\alpha$-(BEDT-TTF)$_2$I$_3$, both theoretically \cite{katayama2006,kino} and experimentally \cite{hirata2016}.
Tilted Dirac cones are commonly classified into type-I, type-II, and type-III: type-I cones are tilted but not overtilted, type-II cones are overtilted, and type-III cones are critically tilted, with the linear dispersion vanishing along one momentum direction.
Type-III Dirac cones have been proposed, based on first-principles calculations, in systems including Zn$_2$In$_2$ \cite{huang2018}, laser-irradiated black phosphorus \cite{Liu}, and Ni$_3$In$_2X_2$ ($X$ = S, Se) \cite{sims}. Type-III Dirac cones have been experimentally observed in artificial photonic systems \cite{Mil}.


Type-III Dirac cones have been shown to appear in simple tight-binding models on a square lattice \cite{mizojpsj}. They exhibit unique transport properties\cite{mizoguchiPRB}  distinct from those of type-I and type-II systems and very weak orbital diamagnetism \cite{Ogata2025} compared with type-I Dirac systems.

In our previous work on critical tilting \cite{KH2025}, we introduced the term ``narrow-sense'' type-III Dirac cone to distinguish the special case in which not only the linear but also all higher-order  terms vanish along the direction connecting the two Dirac points, giving rise to a flat dispersion. In a generic type-III Dirac cone, the linear term vanishes while higher-order terms remain finite. For example, in the three-quarter Dirac cone \cite{KH2017}, the linear term vanishes and a quadratic dispersion remains along one direction, whereas linear dispersions persist along the other three directions.

The narrow-sense type-III Dirac cone is described by the following minimal model:
\begin{equation}
\varepsilon({\pm}, k_x, k_y) = \varepsilon_{\rm F} + \alpha k_y \pm \hbar v_{\rm F} \sqrt{k_x^2 + k_y^2}, \label{model}
\end{equation} 
where $\varepsilon$ is the energy, $\varepsilon_{\rm F}$ is the Fermi energy, $\hbar = h/(2\pi)$, $h$ is Planck's constant, $v_{\rm F}$ is the Fermi velocity, and the parameter $\alpha$ characterizes the tilting of the Dirac cone along the $k_y$ direction, as shown in Fig. \ref{fig1}. 
By using Eq. (\ref{model}), we have revealed a notable crossover in $C$ near the narrow-sense type-III regime, where $C$ changes from $C \propto T^{2}$ below the crossover temperature ($T_{\rm co}$) to $C \propto T^{\frac{1}{2}}$ above $T_{\rm co}$\cite{KH2025}. This crossover is attributed to $\varepsilon$-dependence of the DOS [$D(\varepsilon)$], where $D(\varepsilon) \propto |\varepsilon-\varepsilon_{\rm F}|$ near the Fermi energy ($\varepsilon_{\rm F}$)  (as expected for type-I) and $D(\varepsilon) \propto |\varepsilon-\varepsilon_{\rm F}|^{{-1/2}}$ at energies sufficiently far from $\varepsilon_{\rm F}$ (as expected for narrow-sense type-III).

In a two-dimensional two-site tight-binding model on a square lattice, type-III Dirac cones were shown to arise from degenerate directionally flat bands under a slight modulation of the Hamiltonian \cite{mizojpsj}. In contrast, graphene, a two-dimensional material with a honeycomb lattice structure \cite{wallace}, hosts untilted (type-I) Dirac cones at the K and K$^{\prime}$ points within the nearest-neighbor tight-binding approximation. Recently, tilted Dirac cones in graphene have also been investigated under strain. The strain induces anisotropy in the nearest-neighbor hopping amplitudes and consequently tilts the Dirac cones \cite{Chilla}. In a strained photonic honeycomb lattice with orbital degrees of freedom, a type-III Dirac cone has been shown to emerge from the touching between a flat band and a parabolic band \cite{Mil}.

Although additional hopping processes, such as next-nearest-neighbor terms, can tilt Dirac cones in certain lattice geometries \cite{G2008,kishigi2011}, a narrow-sense type-III Dirac cone with a flat dispersion along one momentum direction has not been demonstrated in a single-orbital honeycomb lattice. In particular, it is interesting to ask whether such a cone can be realized solely through anisotropic next-nearest-neighbor hoppings.

In this work, we demonstrate that narrow-sense type-III Dirac cones can be realized in a single-orbital honeycomb lattice by tuning the next-nearest-neighbor anisotropy. We investigate the resulting band structure, saddle points, density of states, and electronic specific heat across the type-I, narrow-sense type-III, and type-II regimes. Moreover, at the critical point, the additional flat lines in the upper band coincide in energy with the narrow-sense type-III Dirac cones at the Fermi energy, resulting in a pronounced enhancement of the density of states and the electronic specific heat. 

\begin{figure}[bt]
\includegraphics[width=0.43\textwidth]{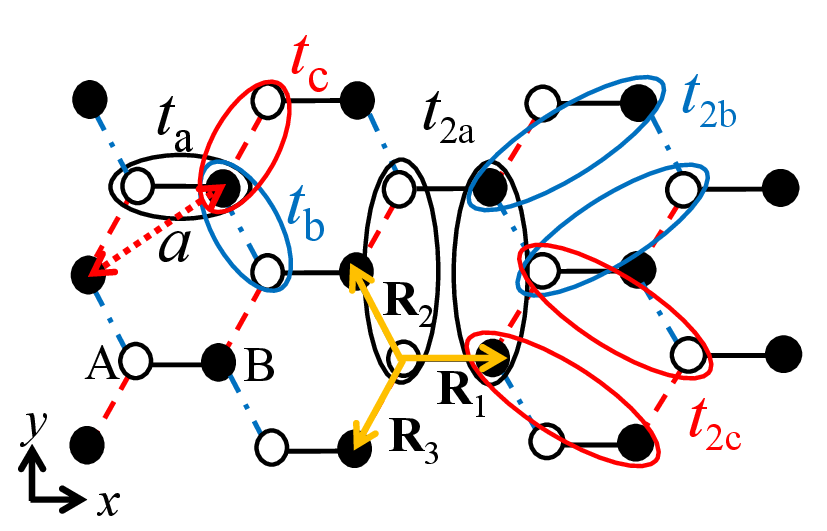}
\caption{(color online).
Honeycomb lattice and transfer integrals.
Filled circles ($\bullet$) and open circles ($\circ$) denote sites
of the A and B sublattices, respectively.
The nearest-neighbor and next-nearest-neighbor transfer integrals
are denoted by $t_a$, $t_b$, and $t_c$ and $t_{2a}$, $t_{2b}$, and $t_{2c}$, respectively.
In this study, we consider $t_a=t_b=t_c=t$ and the anisotropic case
$t_{2a}=0$ with $t_{2b}=t_{2c}=t^{\prime}$. 
}
\label{fig0}
\end{figure}


\begin{figure}[bt]
\begin{flushleft} \hspace{0.5cm}(a) \end{flushleft}\vspace{-0.2cm}
\includegraphics[width=0.41\textwidth]{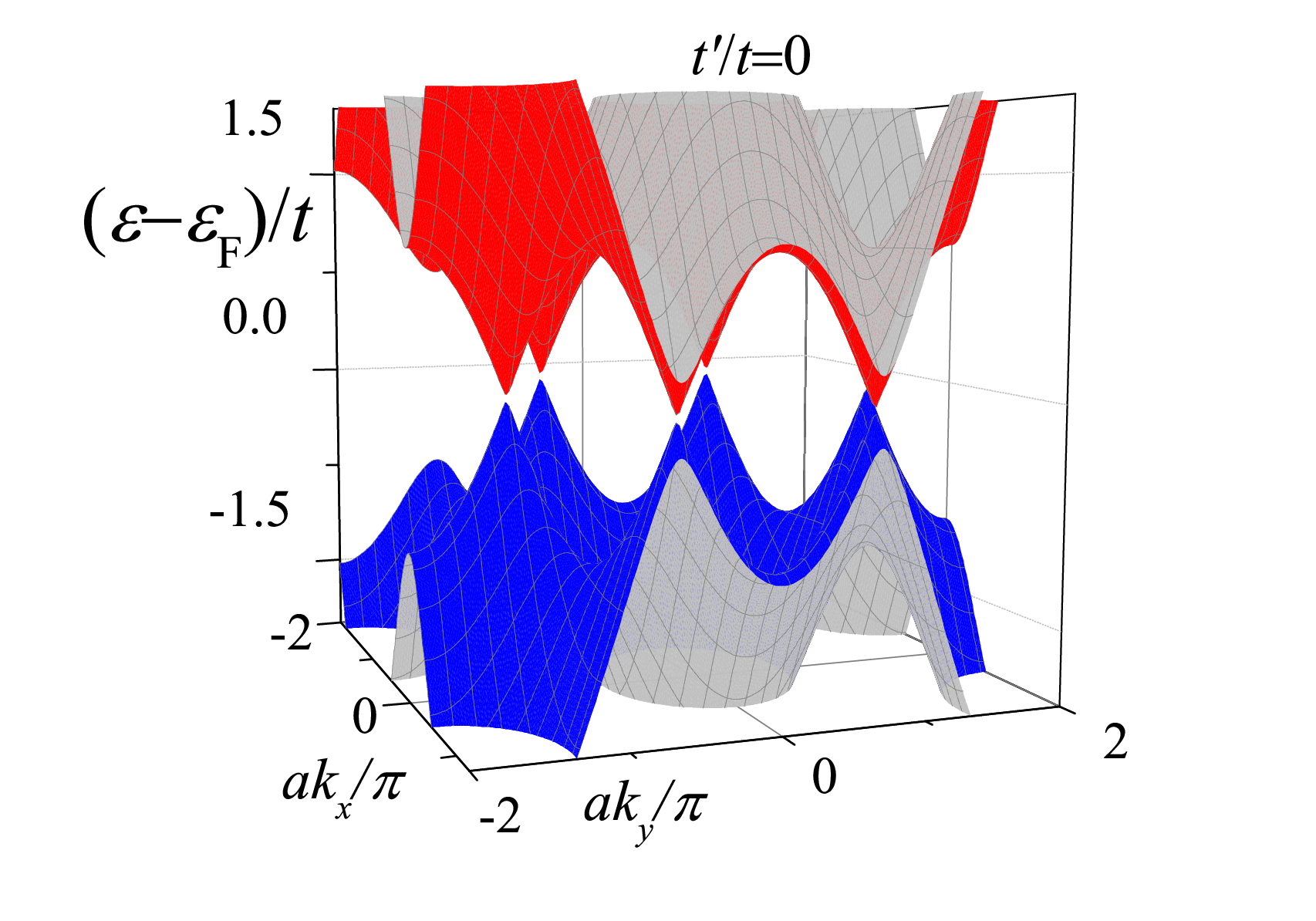}\vspace{-0.2cm}
\begin{flushleft} \hspace{0.5cm}(b) \end{flushleft}\vspace{-0.3cm}
\includegraphics[width=0.41\textwidth]{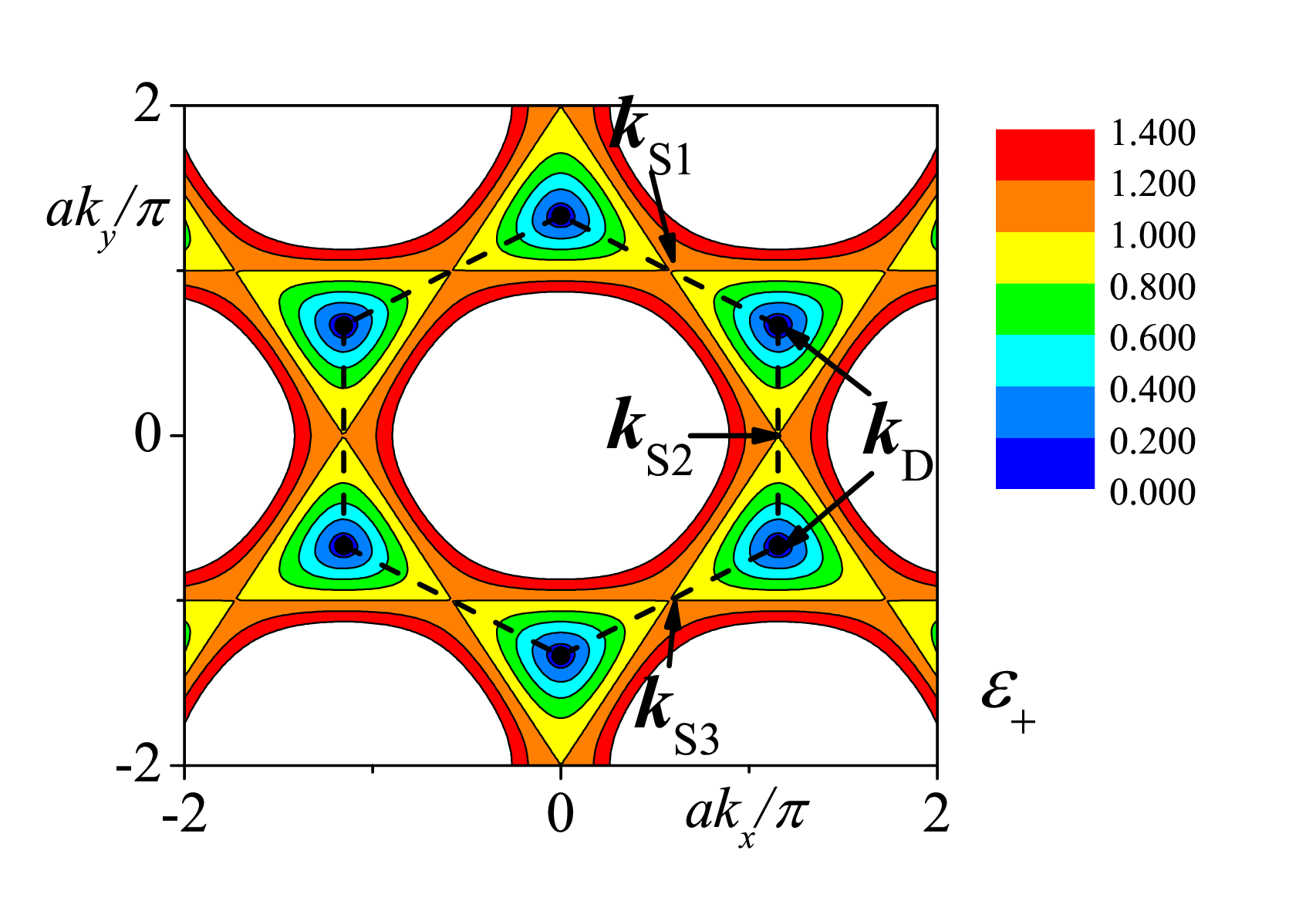}\vspace{-0.2cm}
\begin{flushleft} \hspace{0.5cm}(c) \end{flushleft}\vspace{-0.3cm}
\includegraphics[width=0.41\textwidth]{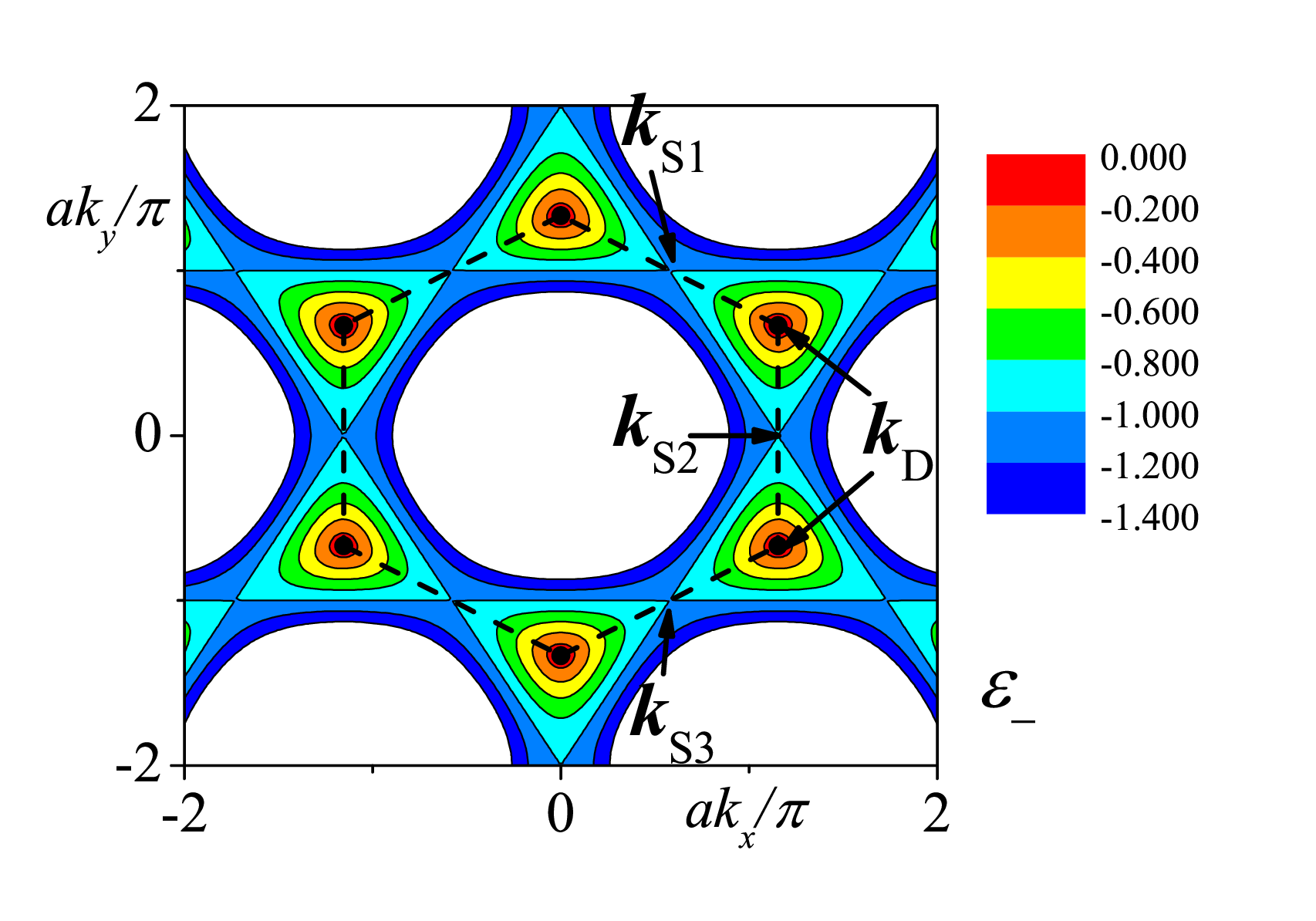}\vspace{-0.1cm}
\caption{(Color online) 
The upper and lower energy bands ($\varepsilon_{+}$ and 
$\varepsilon_{-}$) for $t^{\prime}/t=0$ (a) and contour plots of an upper band (b) and a lower  band (c), where 
the boundary of the first  Brillouin zone are shown by dotted black lines. 
}
\label{fig_band00}
\end{figure}

\begin{figure}[bt]
\begin{flushleft} \hspace{0.5cm}(a) \end{flushleft}\vspace{-0.2cm}
\includegraphics[width=0.43\textwidth]{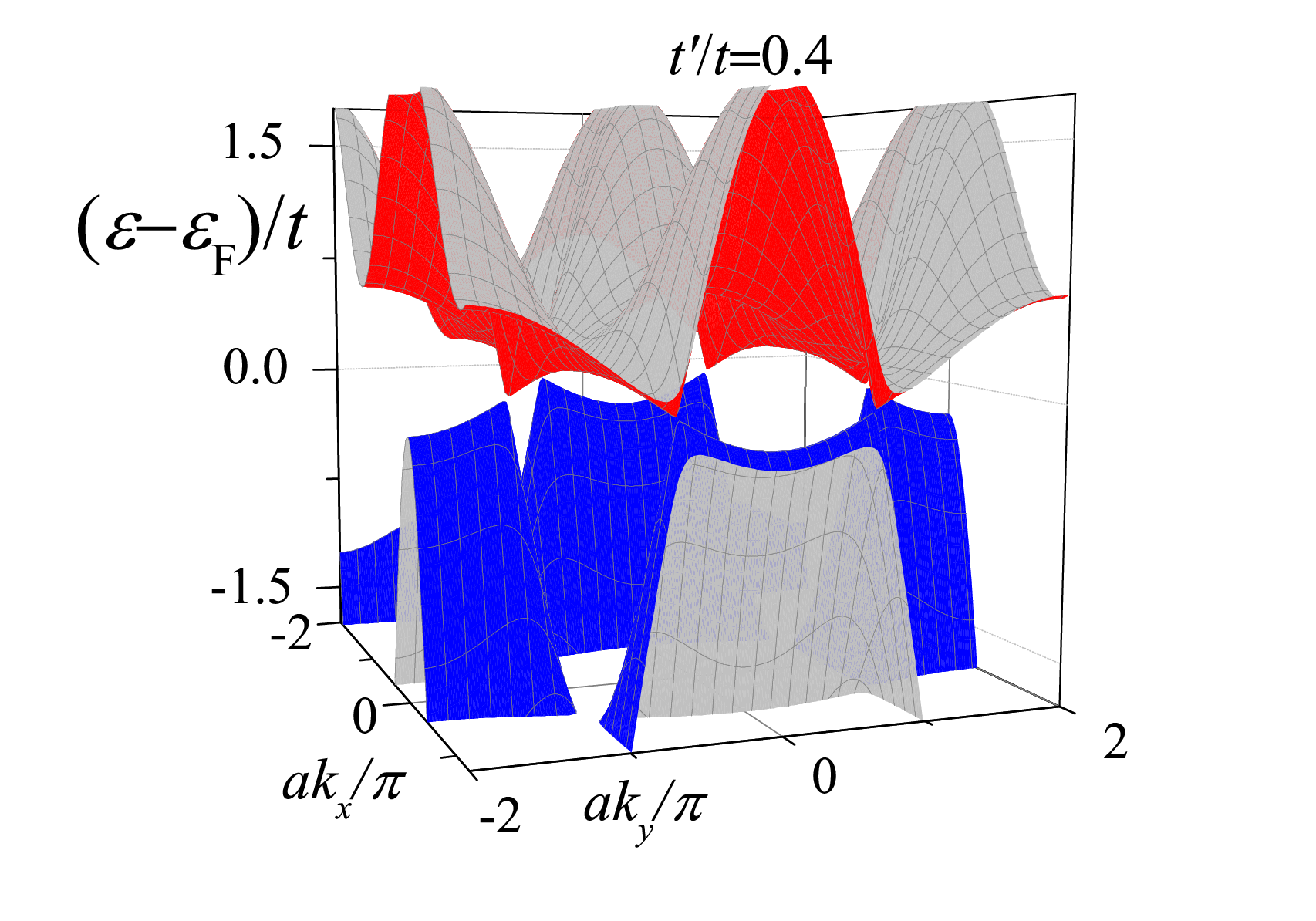}\vspace{-0.2cm}
\begin{flushleft} \hspace{0.5cm}(b) \end{flushleft}\vspace{-0.3cm}
\includegraphics[width=0.43\textwidth]{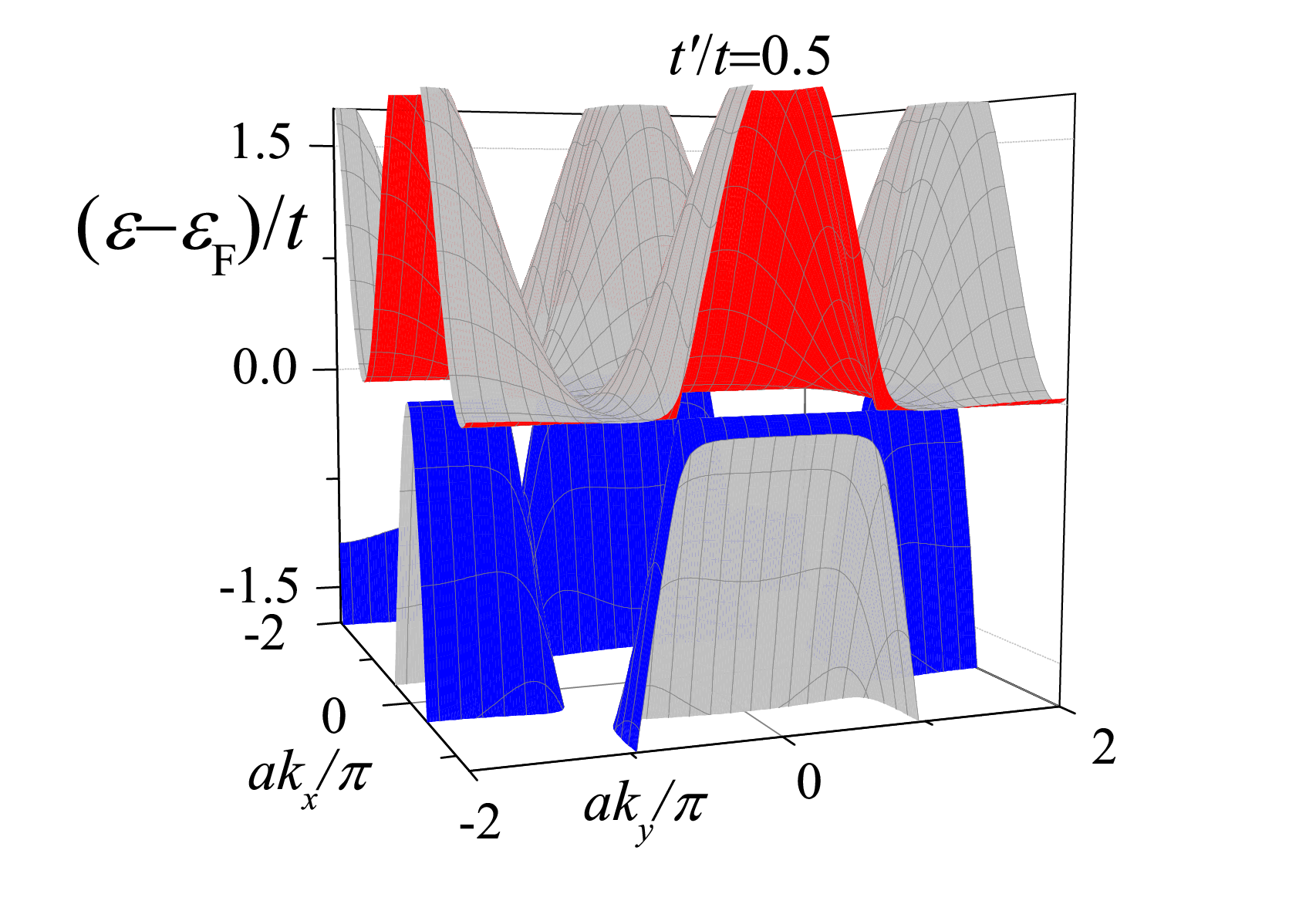}\vspace{-0.2cm}
\begin{flushleft} \hspace{0.5cm}(c) \end{flushleft}\vspace{-0.3cm}
\includegraphics[width=0.43\textwidth]{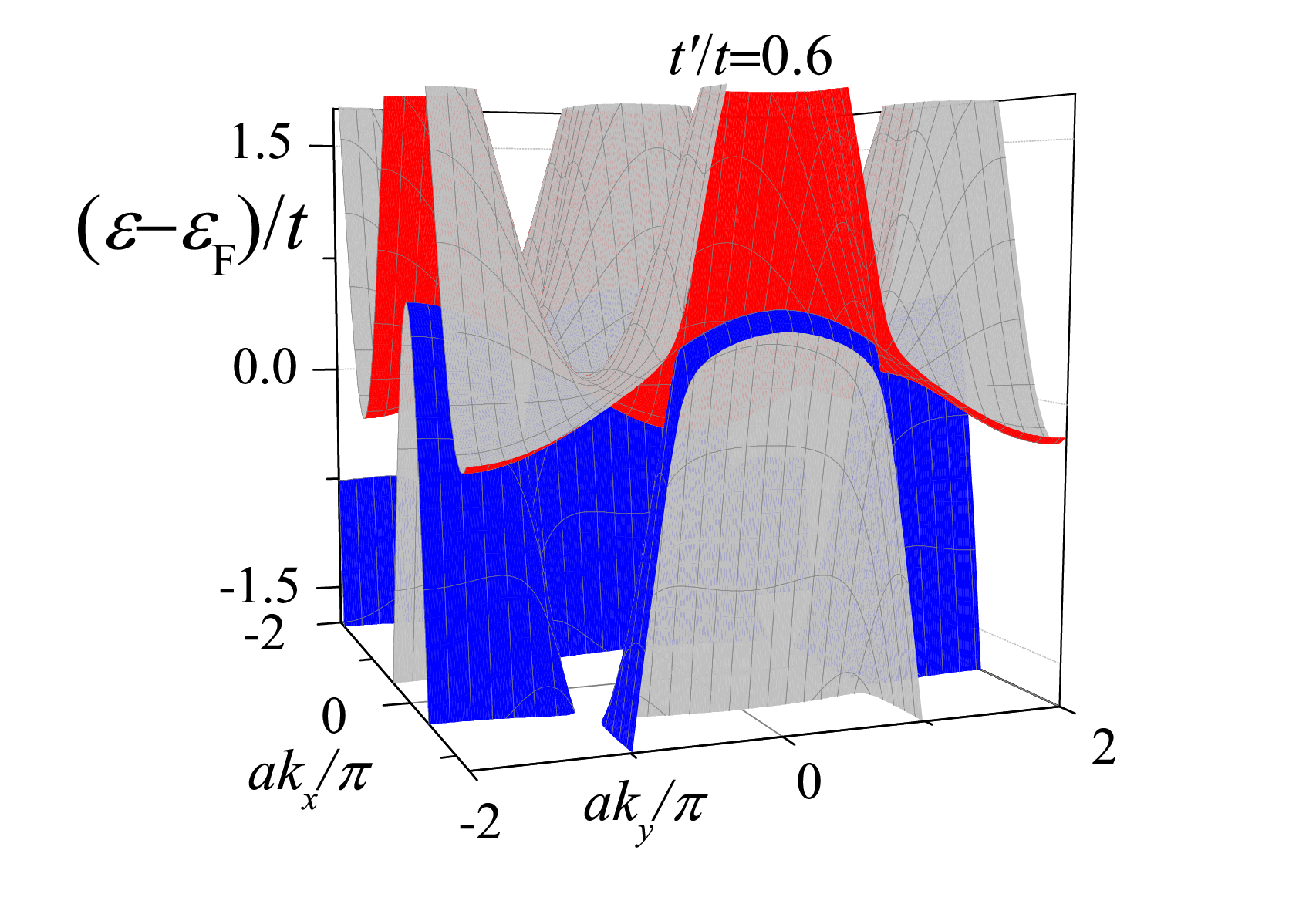}\vspace{-0.1cm}
\caption{(Color online) 
The upper and lower energy bands ($\varepsilon_{+}$ and 
$\varepsilon_{-}$) for $t^{\prime}/t=0.4$ (type-I)  (a), $t^{\prime}/t=0.5$ (narrow-sense type-III) (b), and  $t^{\prime}/t=0.6$ (type-II) (c) when $t_{2a}=0$.
}
\label{fig_band}
\end{figure}

\begin{figure}[bt]
\begin{flushleft} \hspace{0.5cm}(a) \end{flushleft}\vspace{-0.2cm}
\includegraphics[width=0.45\textwidth]{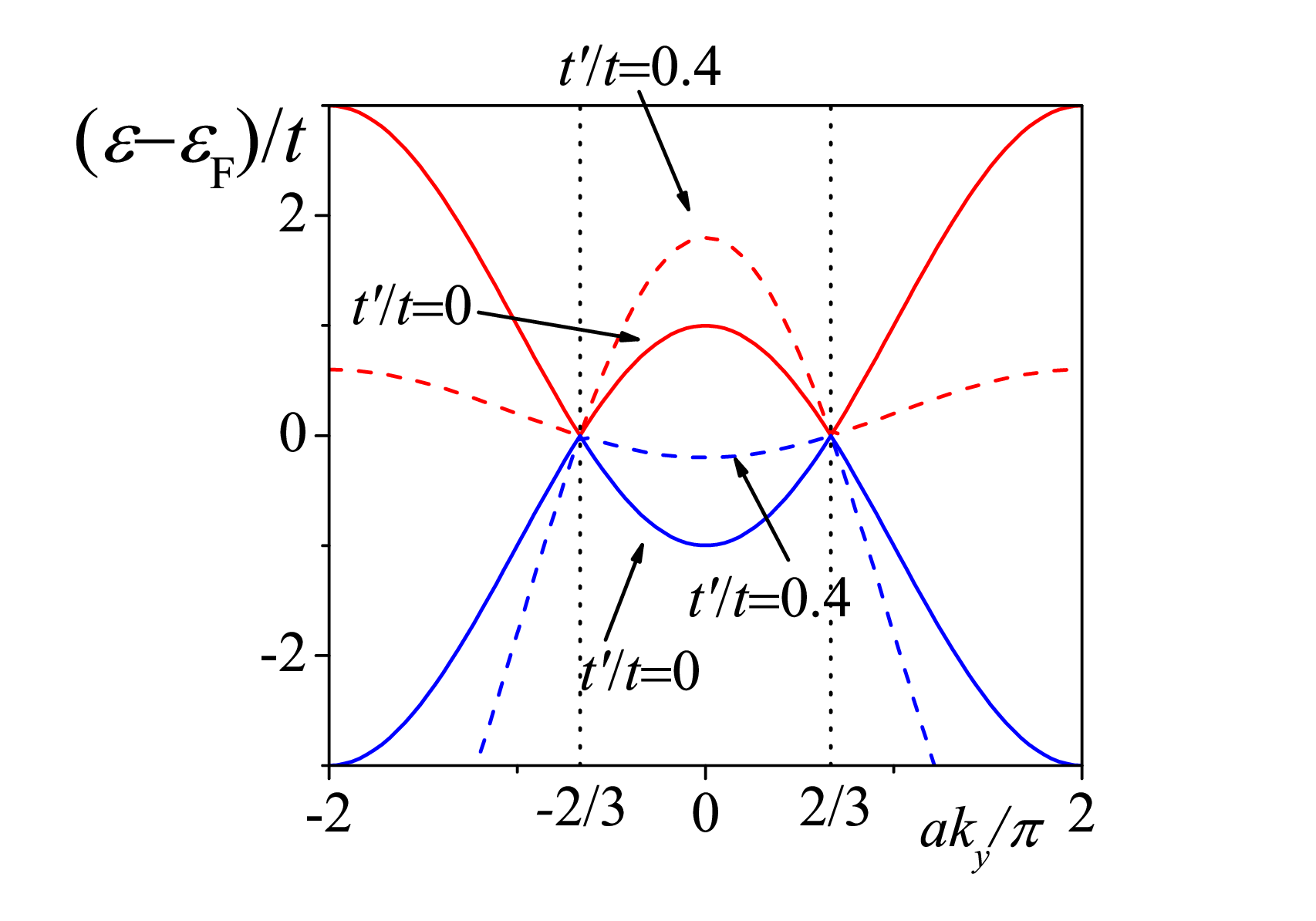}\vspace{-0.2cm}
\begin{flushleft} \hspace{0.5cm}(b) \end{flushleft}\vspace{-0.3cm}
\includegraphics[width=0.45\textwidth]{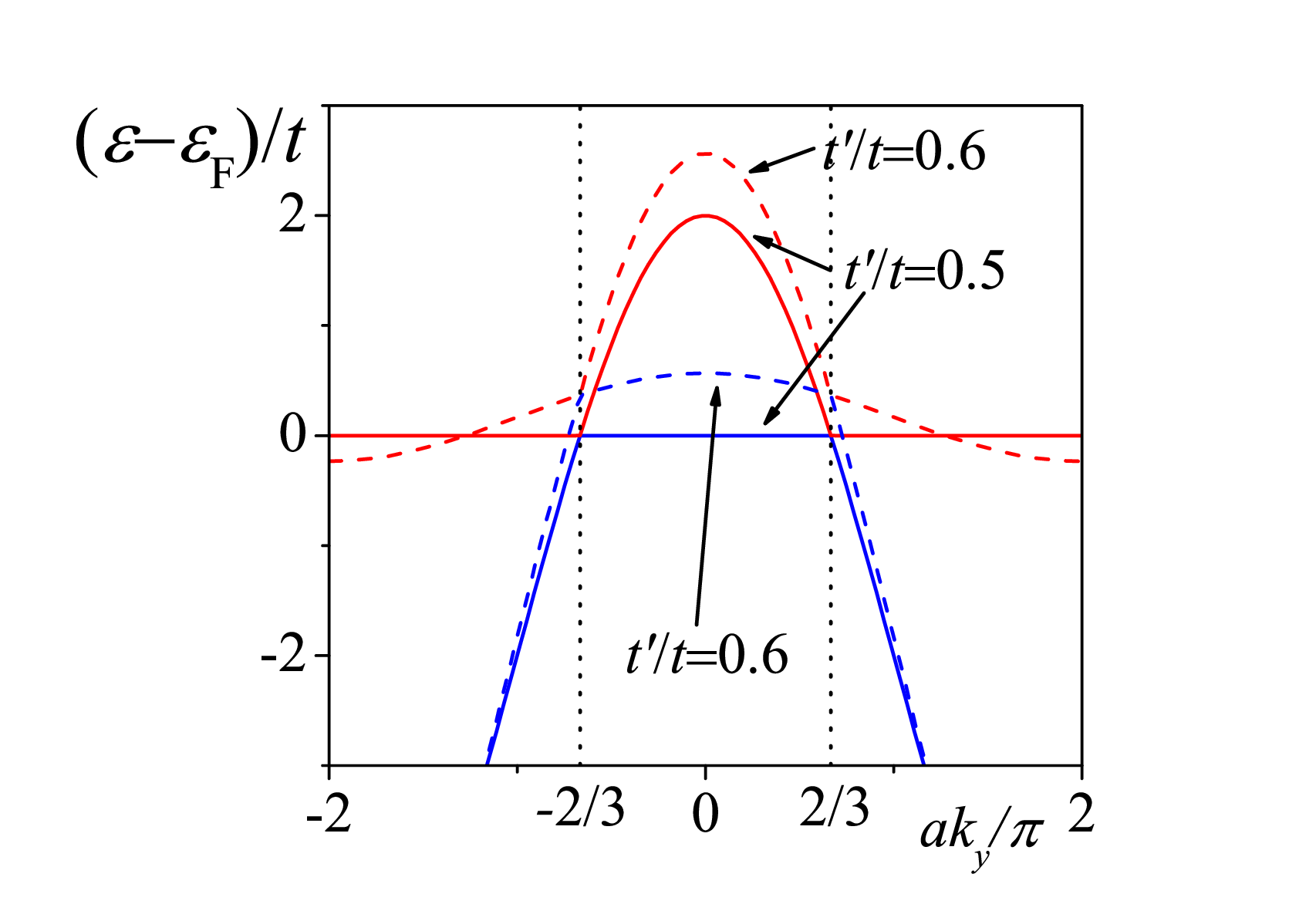}\vspace{-0.2cm}
\caption{(Color online) 
At fixed $k_x = 2\pi/(\sqrt{3}a)$, the upper (red) and lower (blue) energy bands for $t^{\prime}/t=0$ and 0.4 (type-I)  (a) and 0.5 (narrow-sense type-III) and 0.6 (type-II) (b) as a function of $k_y$.  
}
\label{type3}
\end{figure}




\begin{figure}[bt]
\begin{flushleft} 
\end{flushleft}\vspace{-0.2cm}
\includegraphics[width=0.5\textwidth]{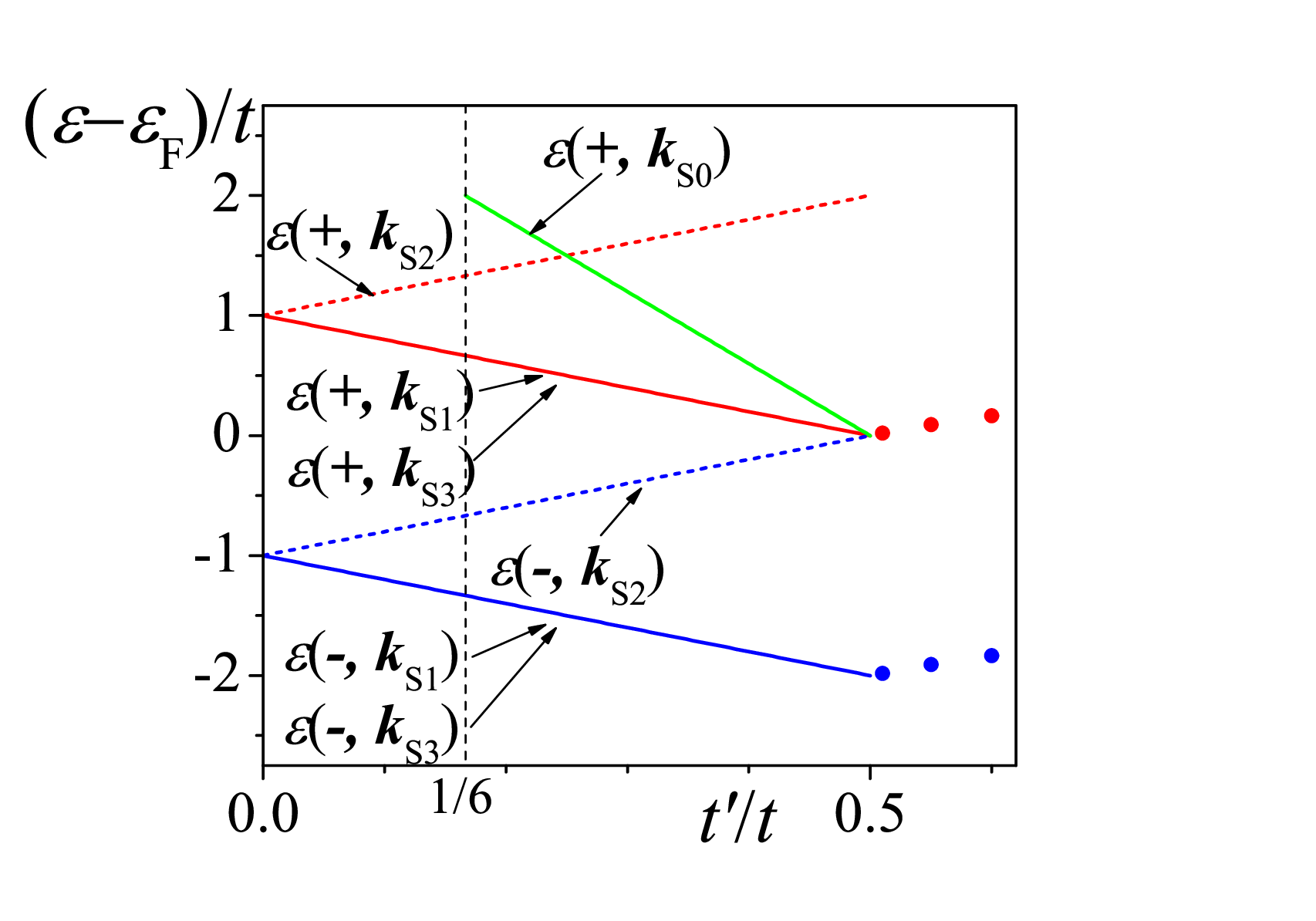}\vspace{-0.2cm}
\caption{(Color online) 
The energy as a function of $t^{\prime}/t$ at the saddle points such as $\varepsilon(\pm, \mathbf{k}_{\rm S1})$, $\varepsilon(\pm, \mathbf{k}_{\rm S2})$,  $\varepsilon(\pm, \mathbf{k}_{\rm S3})$, and $\varepsilon(+, \mathbf{k}_{\rm S0})$. For $1/6<t^{\prime}/t<0.5$,  
$\mathbf{k}_{\rm S0}$ is a saddle point in the upper band. 
The red and blue dots are the energy  of the upper and lower bands at the saddle points,  
$\mathbf{k}_{\rm S1}$ or $\mathbf{k}_{\rm S3}$ for $t^{\prime}/t=0.51, 0.55$, and 0.6. For $0\le t^{\prime}/t <0.5$,  
$\mathbf{k}_{\rm S2}$ is a saddle point in both bands. 
}
\label{fig2_10_2}
\end{figure}

%
\begin{figure}[bt]
\begin{flushleft} \hspace{0.5cm}(a) 
\end{flushleft}\vspace{-0.2cm}
\includegraphics[width=0.45\textwidth]{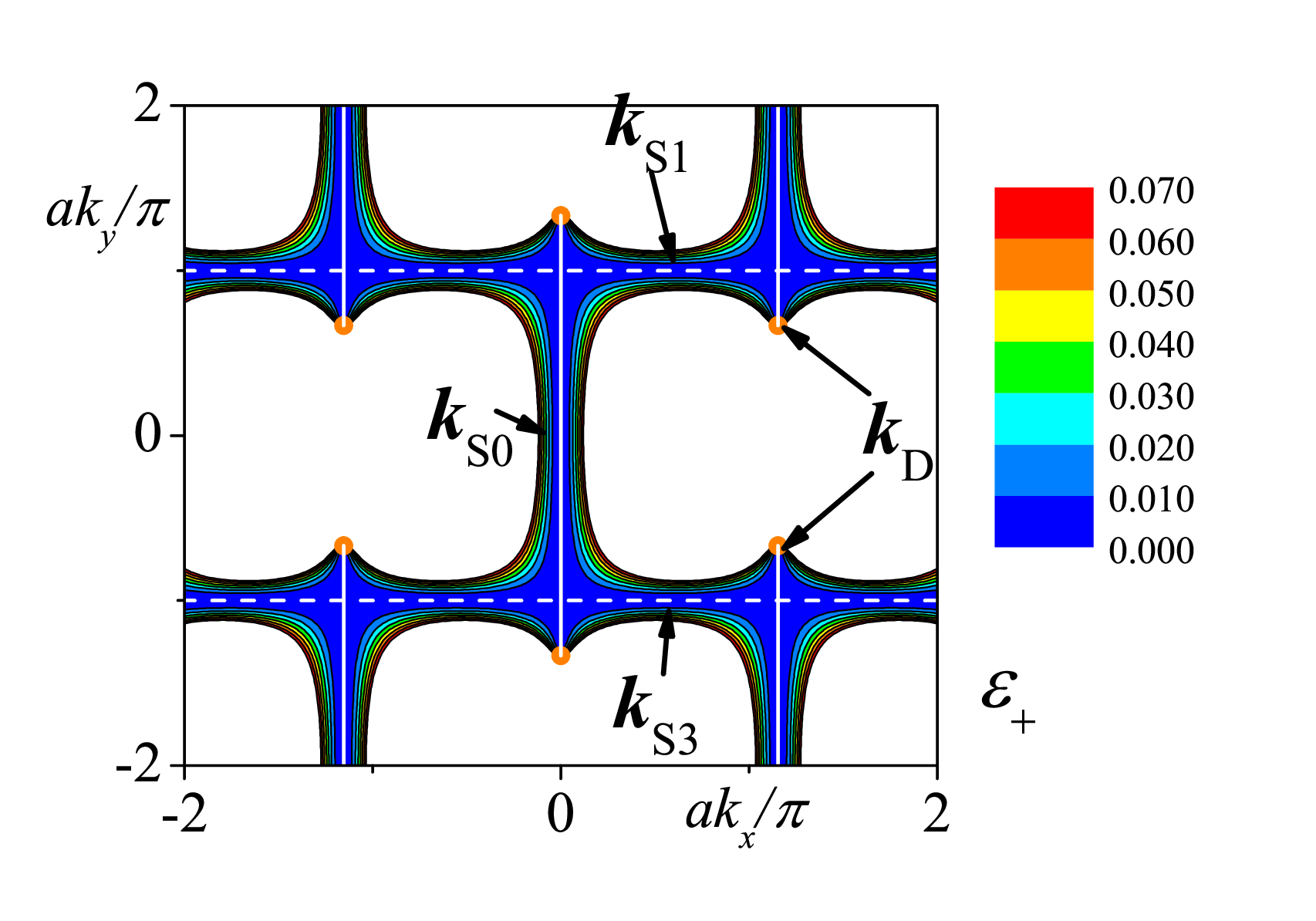}\vspace{-0.2cm}
\begin{flushleft} \hspace{0.5cm}(b) \end{flushleft}\vspace{-0.3cm}
\includegraphics[width=0.45\textwidth]{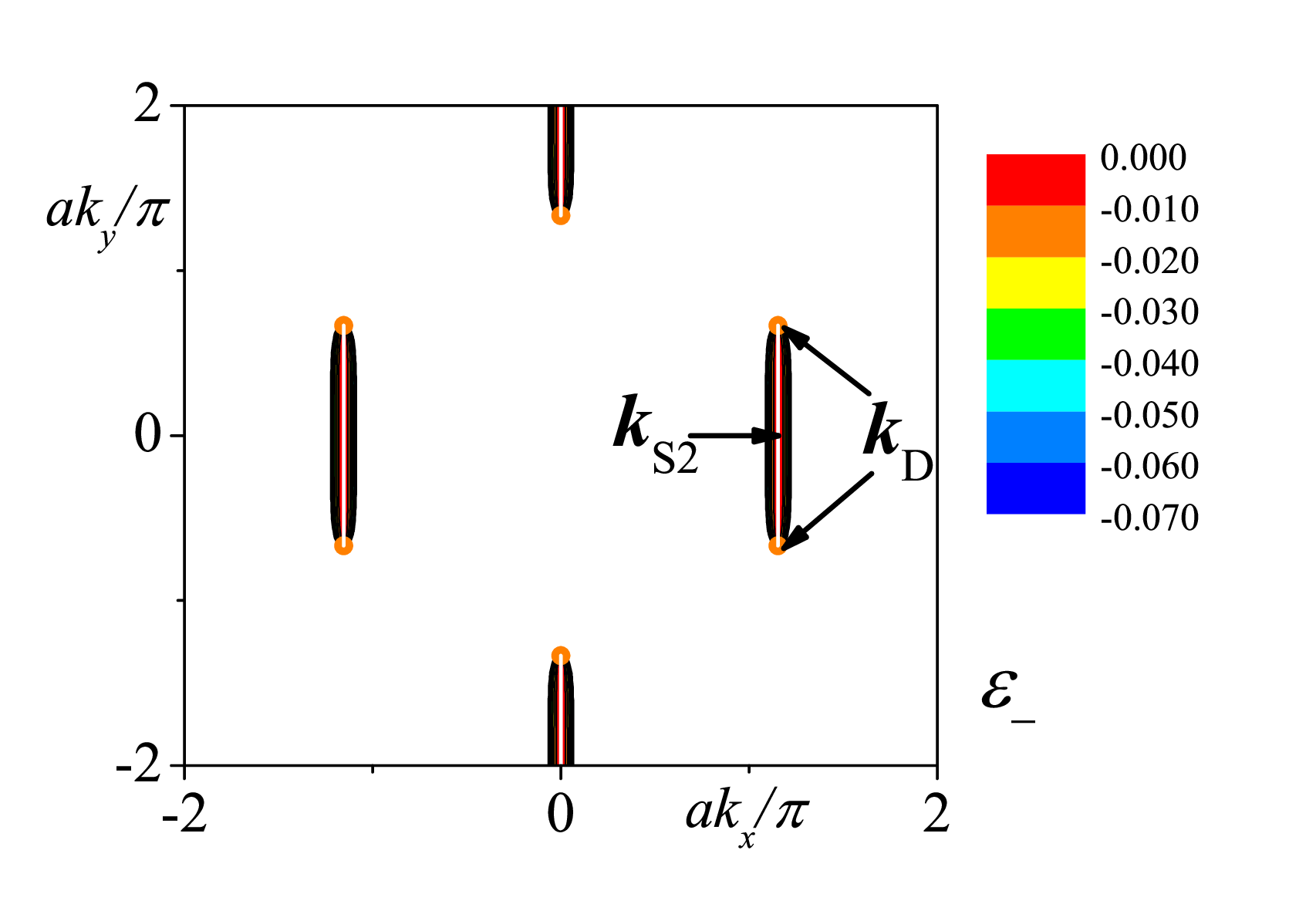}\vspace{-0.2cm}
\caption{(Color online) 
Contour plots of the upper (a) and lower (b) energy bands ($\varepsilon_{+}$ and $\varepsilon_{-}$) for 
Fig. \ref{fig_band} (b) ($t^{\prime}/t = 0.5$), where the Dirac points are indicated  by orange dots. 
In (a) and (b), the white lines represent the flat dispersions at the Fermi energy, 
characteristic of narrow-sense type-III Dirac cones, connecting the two Dirac points in the upper and lower bands.
In (a), the white dotted lines represent the flat lines along the $k_x$ direction at $k_y=\pm \pi/a$ in the upper band. 
}
\label{fig_band_cont2}
\end{figure}

\section{Tight-Binding Model on the Honeycomb Lattice}
\label{EB}

We consider a honeycomb lattice at half filling with nearest- and next-nearest-neighbor
hopping parameters ($t_a$, $t_b$, $t_c$, $t_{2a}$, $t_{2b}$, and $t_{2c}$),
as shown in Fig. \ref{fig0}. For the nearest-neighbor hoppings, we set
$t_a=t_b=t_c=t$, while allowing anisotropy only in the next-nearest-neighbor hoppings.
By varying the anisotropy of the next-nearest-neighbor hoppings, the Dirac cones evolve continuously from type-I to type-II
through the narrow-sense type-III regime. For simplicity, the spin degree of freedom is neglected.

The Hamiltonian is given by
\begin{eqnarray}
\mathscr{H}=\left( 
\begin{array}{cc}
\varepsilon_2({\bf k})
 & F({\bf k})
\\
F^*({\bf k})
&\varepsilon_2({\bf k})
\\
\end{array} 
\right),
\end{eqnarray}
where 
\begin{eqnarray}
F({\bf k})&=&- t_{}{\rm e}^{i{\bf k}\cdot{\bf R}_1} - t_{}{\rm e}^{i{\bf k}\cdot{\bf R}_2} - t_{
}{\rm e}^{i{\bf k}\cdot{\bf R}_3}.\label{f1} \\
\varepsilon_2({\bf k})&=&- 2t_{2c}\cos\bigg[{\bf k}\cdot({\bf R}_1-{\bf R}_2) \bigg]
-2t_{2b}\cos\bigg[{\bf k}\cdot({\bf R}_1-{\bf R}_3)\bigg] \nonumber \\
&-&2t_{2a}\cos\bigg[{\bf k}\cdot({\bf R}_2-{\bf R}_3)\bigg]\label{g1}
\end{eqnarray} 
and 
${\bf R}_1=\big(\frac{\sqrt{3}}{3}a, 0\big),$ ${\bf R}_2=\big(-\frac{\sqrt{3}}{6}a, \frac{1}{2}a\big),$ and 
${\bf R}_3=\big(-\frac{\sqrt{3}}{6}a, -\frac{1}{2}a\big)$, where $a$ is lattice constant, as shown in Fig. \ref{fig0}. 
In this study, we set $t\geq 0$, $t_{2a}\geq 0$,  $t_{2b}\geq 0$, and $t_{2c}\geq 0$. The energy eigenvalues are given by\cite{wallace,kishigi2011} 
\begin{eqnarray}
\varepsilon(\pm,{\bf k})=\varepsilon_2(\mathbf{k})\pm \varepsilon_1(\mathbf{k}),\label{g2}
\end{eqnarray} 
where 
\begin{eqnarray}
\varepsilon_1(\mathbf{k})&=&|F({\bf k})|=
t\bigg[3+ 2\cos( \frac{\sqrt{3}}{2}a k_x - \frac{1}{2}ak_y )\nonumber \\
&+&2\cos( \frac{\sqrt{3}}{2}a k_x + \frac{1}{2}ak_y ) + 2\cos( ak_y )\bigg]^{\frac{1}{2}},\label{c_10}\\
\varepsilon_2(\mathbf{k})&=&-2t_{2c}\cos(\frac{\sqrt{3}}{2}ak_x-\frac{1}{2}ak_y)  \nonumber \\
&&-2t_{2b}\cos(\frac{\sqrt{3}}{2}ak_x+\frac{1}{2}ak_y) -2t_{2a}\cos(ak_y),  \label{c_11} 
\end{eqnarray} 
and $\pm$ signs in $\varepsilon(\pm,{\bf k})$ correspond to upper and lower bands, respectively.

When we neglect the next-nearest-neighbor hoppings (i.e., $t_{2a}=t_{2b}=t_{2c}=0$), 
the energy bands and corresponding contour plots are shown in Fig. \ref{fig_band00}, where 
the Dirac points  ($\mathbf{k}_{\rm D}$)  are located at the K and K' points of the Brillouin zone. 
Even in the presence of next-nearest-neighbor hoppings, the positions of the Dirac points remain unchanged, whereas the Dirac cones become tilted\cite{kishigi2011}.
The three $M$ points are given by \cite{gill}
\begin{eqnarray}
{\bf k}_{\rm S1}&=&(\pi /(\sqrt{3}a), \pi/a),\\
{\bf k}_{\rm S2}&=&(2\pi /(\sqrt{3}a), 0),\\
{\bf k}_{\rm S3}&=&(\pi /(\sqrt{3}a), -\pi/a).
\end{eqnarray} 


\section{Anisotropic Next-Nearest-Neighbor Hopping}


\subsection{Emergence of narrow-sense type-III Dirac cones}

We consider an anisotropic next-nearest-neighbor hopping
with $t_{2a}=0$ and $t_{2b}=t_{2c}=t^{\prime}$,
and restrict the parameter range to
$0\leq t^{\prime}/t\leq 0.6$. 
The contribution from the next-nearest-neighbor hopping is then given by
\begin{eqnarray}
\varepsilon_2(\mathbf{k}) = -2t^{\prime}\bigg[
\cos(\tfrac{\sqrt{3}}{2}ak_x - \tfrac{1}{2}ak_y)
+\cos (\tfrac{\sqrt{3}}{2}ak_x + \tfrac{1}{2}ak_y)
\bigg].
\label{c_14}
\end{eqnarray}

For $0 \le t^{\prime}/t \le 0.5$ the Fermi energy coincides with the Dirac-point energy, yielding
\begin{eqnarray}
\varepsilon_{\rm F} = \varepsilon_2(\mathbf{k}_{\rm D}) = 2t^{\prime}.
\label{ef}
\end{eqnarray}
For  $t^{\prime}/t > 0.5$, since an electron pocket and a hole pocket overlap, the Fermi energy must be determined numerically.
For example, for  $t^{\prime}/t= 0.6$, we obtain  
$\varepsilon_{\rm F} \simeq 0.833t$. 

By varying $t^{\prime}/t$, the Dirac cones become tilted along the $k_y$ direction. For $0 \le t^{\prime}/t < 0.5$, $t^{\prime}/t=0.5$, and $t^{\prime}/t>0.5$, they are classified as type-I, narrow-sense type-III, and type-II Dirac cones, respectively, as shown in Fig.~\ref{fig_band}. We now verify this classification analytically.

By expanding $\varepsilon(\mathbf{k})$ around the Dirac point $\mathbf{k}_{\rm D}= (2\pi/(\sqrt{3}a), 2\pi/(3a))$, we obtain
\begin{eqnarray}
\varepsilon(\pm, \mathbf{k}) &\simeq& 2t^{\prime}
- \sqrt{3} t^{\prime} \left(ak_y - \tfrac{2\pi}{3}\right) \nonumber \\
&\pm& \frac{\sqrt{3}}{2}t
\sqrt{\left(ak_x - \tfrac{2\pi}{\sqrt{3}}\right)^2 
+ \left(ak_y - \tfrac{2\pi}{3}\right)^2}.
\label{e_ns}
\end{eqnarray}
Equation~(\ref{e_ns}) describes the Dirac cones near $\mathbf{k}_{\rm D}$. They are classified as type-I, type-II, and narrow-sense type-III for
$t^{\prime}/t < 0.5$, $> 0.5$, and $=0.5$, respectively.

The energy dispersion along the $k_y$ direction at fixed $k_x = 2\pi/(\sqrt{3}a)$ is given by
\begin{eqnarray}
\varepsilon(\pm, \tfrac{2\pi}{\sqrt{3}a}, k_y)
= 4t^{\prime}\cos(\tfrac{1}{2}ak_y)
\pm t \left|1 - 2\cos (\tfrac{1}{2}ak_y)\right|, 
\label{ky_d}
\end{eqnarray}
which is shown in Fig. \ref{type3}, where for $t^{\prime}/t = 0.5$ flat dispersions along the $k_y$ direction between the two Dirac points are observed 
in an upper band at $-2\pi /a< k_y\le -2\pi /(3a)$ and $2\pi /(3a)< k_y\le 2\pi /a$, 
in a lower band  at $-2\pi /(3a)< k_y\leq 2\pi /(3a)$. In fact, at $t^{\prime}/t = 0.5$, Eq. (\ref{ky_d}) becomes piecewise constant:
\begin{align}
&\varepsilon(+, \tfrac{2 \pi}{\sqrt{3}a}, k_y) - \varepsilon_{\rm F} \nonumber\\
&=
\begin{cases}
0 & \left(-\tfrac{2\pi}{a}<k_y\leq -\tfrac{2\pi}{3a},\ \tfrac{2\pi}{3a}<k_y\leq\tfrac{2\pi}{a}\right), \\
4t \cos(\tfrac{1}{2}ak_y) -2t & \left(-\tfrac{2\pi}{3a}<k_y\leq\tfrac{2\pi}{3a}\right),
\end{cases}
\label{ky+dep}
\end{align}
\begin{align}
&\varepsilon(-, \tfrac{2 \pi}{\sqrt{3}a}, k_y) - \varepsilon_{\rm F} \nonumber\\
&=
\begin{cases}
4t \cos(\tfrac{1}{2}ak_y) - 2t & \left(-\tfrac{2\pi}{a}<k_y\leq-\tfrac{2\pi}{3a},\ \tfrac{2\pi}{3a}<k_y\leq\tfrac{2\pi}{a}\right), \\
0 & \left(-\tfrac{2\pi}{3a}< k_y\leq\tfrac{2\pi}{3a}\right).
\end{cases}
\label{ky+dep2}
\end{align}
Thus, the energy becomes independent of $k_y$ over finite intervals, giving rise to flat dispersions at
$(\varepsilon-\varepsilon_{\rm F})/t=0$. This confirms the realization of a narrow-sense type-III Dirac cone.

\subsection{Saddle points}

Details of the saddle-point analysis are given in the
Supplemental Material\cite{SM}.

In addition to the saddle points at
$\mathbf{k}_{\rm S1}$, $\mathbf{k}_{\rm S2}$, and $\mathbf{k}_{\rm S3}$,
we find that the $\Gamma$ point,
\begin{eqnarray}
\mathbf{k}_{\rm S0}=(0,0),
\end{eqnarray}
becomes a saddle point in the upper band for
$1/6<t^{\prime}/t<0.5$, while it is always a local minimum
in the lower band.

For $0\leq t^{\prime}/t<0.5$ and $t^{\prime}/t>0.5$,
$\mathbf{k}_{\rm S1}$ is a saddle point in both bands.
At $t^{\prime}/t=0.5$, $\mathbf{k}_{\rm S1}$ belongs to a
degenerate set of local minima and is not a saddle point in the upper band, 
whereas $\mathbf{k}_{\rm S1}$ remains a saddle point in the lower band. Moreover, $\mathbf{k}_{\rm S1}$ and
$\mathbf{k}_{\rm S3}$ have the same saddle-point character and
are degenerate in energy:
\begin{eqnarray}
\varepsilon(\pm,\mathbf{k}_{\rm S1})
=
\varepsilon(\pm,\mathbf{k}_{\rm S3}).
\end{eqnarray}

For $0\leq t^{\prime}/t<0.5$, $\mathbf{k}_{\rm S2}$ is a saddle point  in both bands. 
At $t^{\prime}/t=0.5$, $\mathbf{k}_{\rm S2}$ is a local maximum in the upper band, while in the lower band it belongs to a
degenerate set of local maxima. Thus, $\mathbf{k}_{\rm S2}$ is not a saddle point in either band. For $t^{\prime}/t>0.5$,
$\mathbf{k}_{\rm S2}$ is a local maximum in both bands and is no longer a saddle point.

The energies of the stationary points are summarized below, together with the parameter ranges in which they are saddle points. 
\begin{eqnarray}
\varepsilon(+, \mathbf{k}_{\rm S0}) &=& 3t-4t^{\prime},\qquad  1/6< t^{\prime}/t<0.5,
\label{ks3}\\
\varepsilon(-, \mathbf{k}_{\rm S1}) &=& \varepsilon(-, \mathbf{k}_{\rm S3}) =-t, \qquad t^{\prime}/t\geq 0,
\label{ks2_}\\
\varepsilon(+, \mathbf{k}_{\rm S1}) &=& \varepsilon(+, \mathbf{k}_{\rm S3}) =t, \qquad  t^{\prime}/t\neq 0.5,
\label{ks2}\\
\varepsilon(\pm, \mathbf{k}_{\rm S2}) &=& \pm t+4t^{\prime},\qquad   0\leq t^{\prime}/t<0.5.
\label{ks1} 
\end{eqnarray}
The energies listed above are shown in Fig.~\ref{fig2_10_2}.

\subsection{Flat lines along the $k_x$ direction at $k_y=\pm\pi/a$}

A characteristic feature of the present model is the existence of flat lines along the $k_x$ direction at $k_y=\pm\pi/a$. 
These flat lines are indicated by white dotted lines in Fig.~\ref{fig_band_cont2} (a). We now clarify their origin.

Substituting $k_y=\pm\pi/a$ into Eqs.~(\ref{c_10}) and (\ref{c_14}), we obtain
\begin{eqnarray}
\varepsilon_1(k_x,\pm\tfrac{\pi}{a}) &=& t,\\
\varepsilon_2(k_x,\pm\tfrac{\pi}{a}) &=& 0.
\label{ef_aN_0}
\end{eqnarray}
Consequently, the energy bands become
\begin{eqnarray}
\varepsilon(\pm,k_x,\pm\tfrac{\pi}{a})=\pm t,
\end{eqnarray}
which are independent of $k_x$, giving rise to flat lines. This feature is characteristic of the present model and is absent in the square lattice model studied previously \cite{Ogata2025}.

For $0\leq t^{\prime}/t\leq0.5$,
\begin{eqnarray}
\varepsilon(\pm,k_x,\pm\tfrac{\pi}{a})-\varepsilon_{\rm F}
=\pm t-2t^{\prime}.
\label{ef_aN2}
\end{eqnarray}
Thus, at $t^{\prime}/t=0.5$, $\varepsilon(+,k_x,\pm\tfrac{\pi}{a})-\varepsilon_{\rm F}=0$, whereas $\varepsilon(-,k_x,\pm\tfrac{\pi}{a})-\varepsilon_{\rm F}=-2t$. The flat lines in the upper band therefore lie at the Fermi energy, as shown in Fig.~\ref{fig_band_cont2} (a).

The energy contour of the upper band in Fig.~\ref{fig_band_cont2}(a) also shows a flat dispersion in the vicinity of $k_y=\pm\pi/a$. To examine this feature, we expand 
$\varepsilon(+,\mathbf{k})$ around $k_y=\pi/a$. 
The upper-band dispersion near the flat line is given by
\begin{eqnarray}
\varepsilon(+,\mathbf{k})
\simeq&t
+\left(2t'-t\right)
\cos(\frac{\sqrt{3}}{2}ak_x)\left(ak_y-\pi\right)
\nonumber \\
&+\frac{1}{2}t\sin^2(\frac{\sqrt{3}}{2}ak_x)
\left(ak_y-\pi\right)^2.
\end{eqnarray}
At the critical value $t'/t=0.5$, the linear term in the momentum displacement perpendicular to the flat line vanishes. The leading transverse dispersion is therefore quadratic for generic $k_x$ along the flat line. This suppression of the $k_y$ dispersion, together with the flat dispersion along the $k_x$ direction, contributes to the strong enhancement of the DOS above the Fermi energy, as discussed below.

For $t'/t\neq0.5$, the linear term in the transverse dispersion is generally nonzero, even though the energy remains independent of $k_x$ exactly on the lines ($k_y=\pm\pi/a$). This distinction is important for understanding the DOS near the Fermi energy.

\begin{figure}[bt]
\begin{flushleft} 
\hspace{0.5cm}(a) \end{flushleft}\vspace{-0.2cm}
\includegraphics[width=0.43\textwidth]{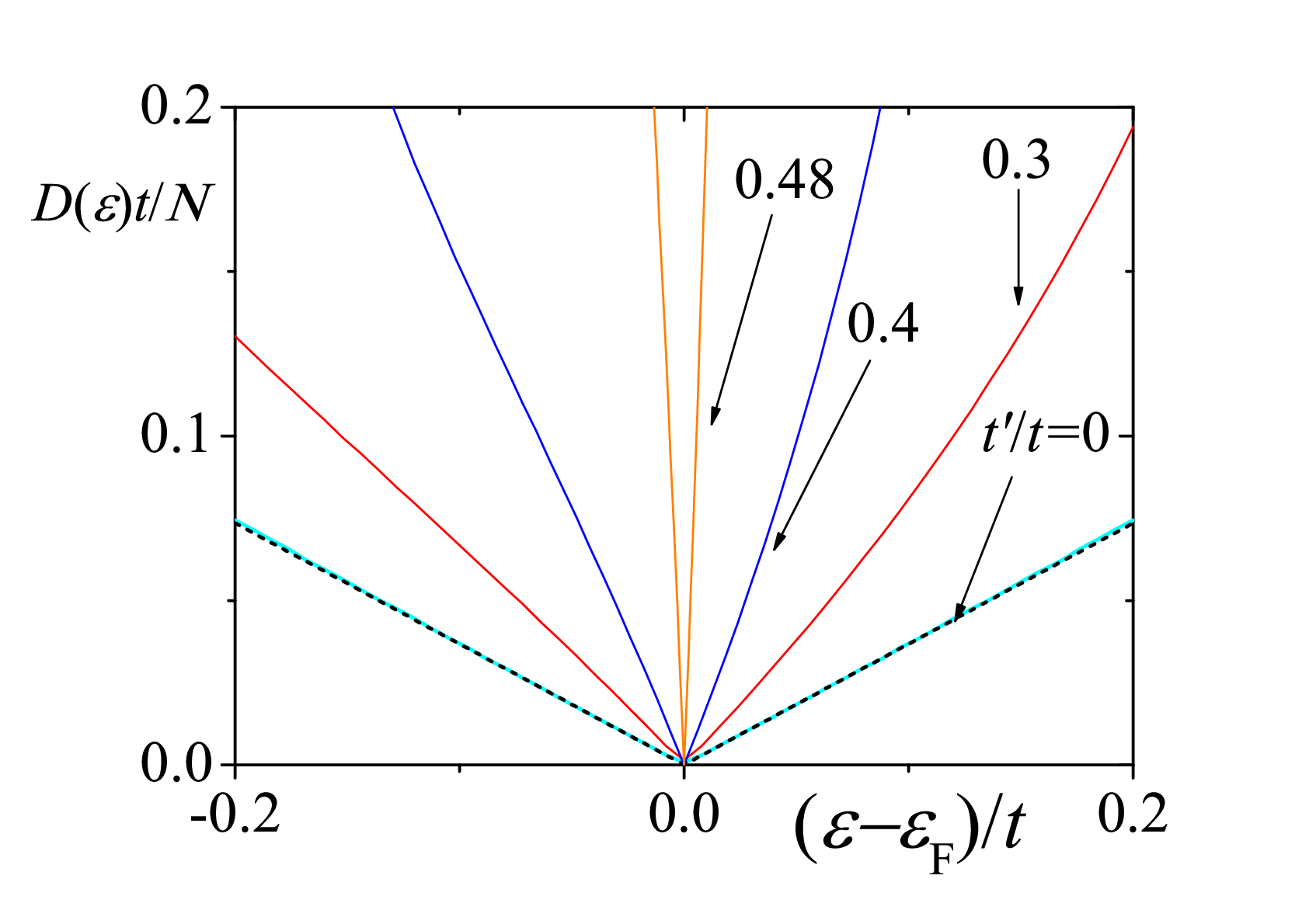}\vspace{-0.2cm}
\begin{flushleft} \hspace{0.5cm}(b) \end{flushleft}\vspace{-0.3cm}
\includegraphics[width=0.43\textwidth]{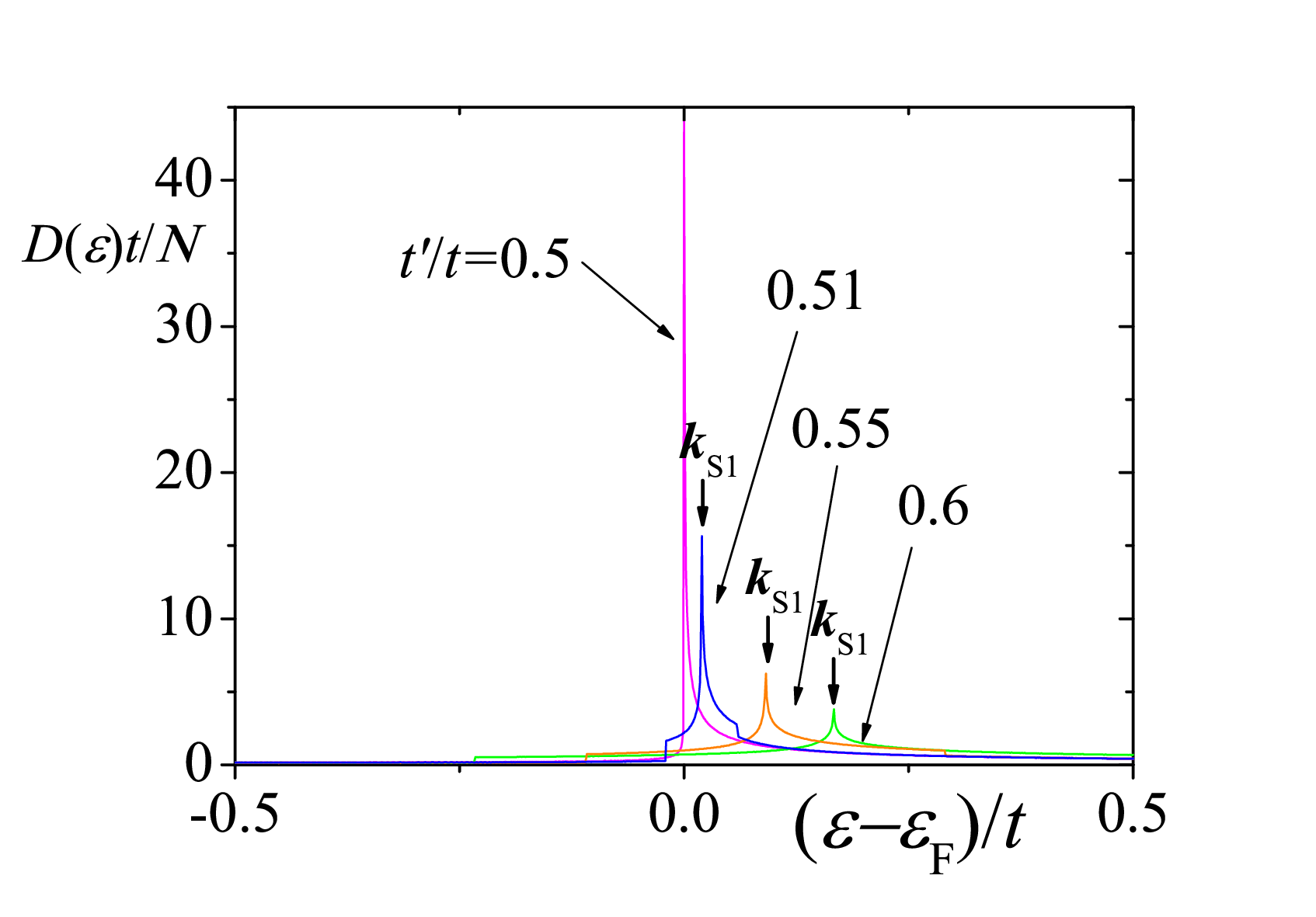}\vspace{-0.2cm}
\begin{flushleft} \hspace{0.5cm}(c) \end{flushleft}\vspace{-0.3cm}
\includegraphics[width=0.43\textwidth]{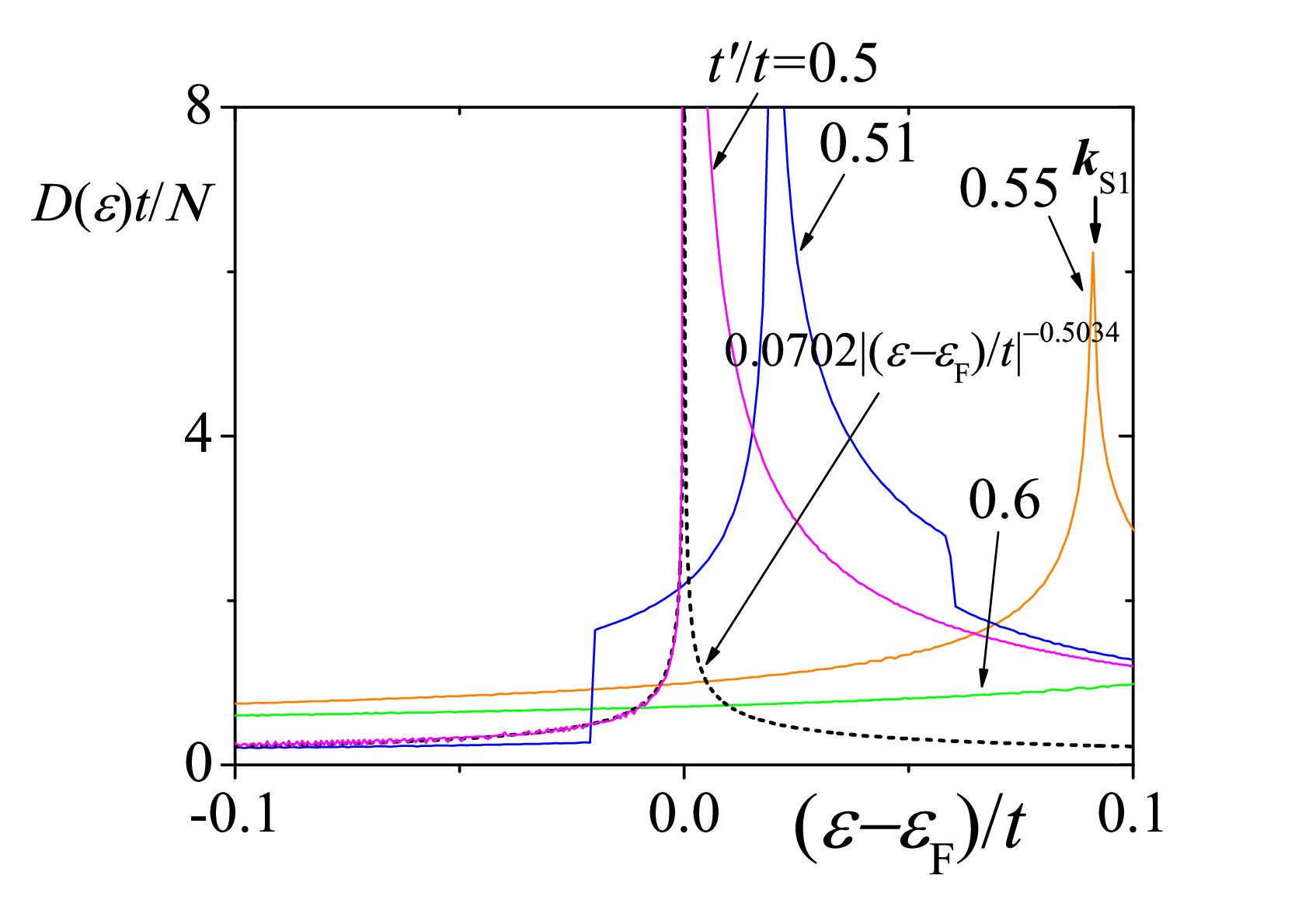}\vspace{-0.2cm}
\caption{(Color online) 
The DOS at different $t^{\prime}/t$ values.
(a) shows the type-I cases with $t^{\prime}/t=0, 0.3, 0.4,$ and $0.48$. 
(b) shows the narrow-sense type-III case with $t^{\prime}/t=0.5$
and the type-II cases with $t^{\prime}/t=0.51, 0.55,$ and $0.6$.
For $t^{\prime}/t=0.51, 0.55,$ and $0.6$,
small peaks appear at
$(\varepsilon(+, \mathbf{k}_{\rm S1})-\varepsilon_{\rm F})/t
=0.020, 0.091,$ and $0.167$, respectively.
(c) is an enlarged view of (b). 
}
\label{fig_dos}
\end{figure}

\subsection{DOS in type-I, narrow-sense type-III, and type-II}

For Eq.~(\ref{model}) with $\alpha=0$, taking the first Brillouin zone of the honeycomb lattice into account, 
the DOS near the Fermi energy is approximately given by\cite{KH2025}
\begin{eqnarray}
D(\varepsilon)=
\frac{2\sqrt{3}N}{3\pi t}
|(\varepsilon-\varepsilon_{\rm F})/t|,
\qquad
\mbox{(type-I)},
\label{d_0_1_1}
\end{eqnarray}
where $N$ denotes the number of $\mathbf{k}$ points sampled in the first Brillouin zone. 
The numerically obtained  DOS in the present model is shown in Fig.~\ref{fig_dos}. For $0\le t^{\prime}/t<0.5$, 
the Dirac cone remains of type-I, and therefore the DOS exhibits the characteristic linear dependence 
$D(\varepsilon)\propto |\varepsilon-\varepsilon_{\rm F}|$ near the Fermi energy. 
Equation~(\ref{d_0_1_1}) (black dotted line) agrees well with the numerical result (sky-blue line) for $t^{\prime}/t=0$, as shown in Fig.~\ref{fig_dos}(a). 
For  $t^{\prime}/t>0$, the linear energy dependence is preserved, although the prefactor is modified by the tilt of the Dirac cone.

In the Supplemental Material\cite{SM}, we show the DOS for $t'/t=0$, 0.3, 0.4, and 0.48, illustrating the van Hove singularities associated with the saddle points ($\mathbf{k}_{\rm S1}$, $\mathbf{k}_{\rm S2}$, and $\mathbf{k}_{\rm S3}$). In particular, the saddle point at the $\Gamma$ point, 
$\mathbf{k}_{\rm S0}$, in the upper band for $1/6<t'/t<0.5$ gives rise to an additional van Hove singularity. As $t'/t$ approaches 0.5 from below,  the energies of van Hove singularities due to the upper-band saddle points ($\mathbf{k}_{\rm S0}$, $\mathbf{k}_{\rm S1}$, and $\mathbf{k}_{\rm S3}$) and the lower-band saddle point ($\mathbf{k}_{\rm S2}$) move toward the Fermi energy, as shown in Fig. \ref{fig2_10_2}. At the critical value $t'/t=0.5$, these stationary points become part of degenerate sets of stationary points along flat lines. Specifically, $\mathbf{k}_{\rm S1}$ and $\mathbf{k}_{\rm S3}$ lie on the additional flat lines at $k_y=\pm\pi/a$, whereas $\mathbf{k}_{\rm S0}$ and $\mathbf{k}_{\rm S2}$ lie on the flat lines connecting the two Dirac points at $k_x=0$ and $2\pi/(\sqrt{3}a)$, respectively, as shown in Fig.~\ref{fig_band_cont2}. 


At $t^{\prime}/t=0.5$ (narrow-sense type-III), a prominent peak in the DOS emerges at $\varepsilon=\varepsilon_{\rm F}$, 
as shown in Fig.~\ref{fig_dos}(b), with an enlarged view in Fig.~\ref{fig_dos}(c).
For the lower band 
($\varepsilon<\varepsilon_{\rm F}$), the DOS in the range of 
$-10^{-2}\leq (\varepsilon-\varepsilon_{\rm F})/t\leq -4.0\times10^{-4}$ 
is well fitted by
\begin{eqnarray}
D(\varepsilon)
=\frac{0.0702N}{t}
|(\varepsilon-\varepsilon_{\rm F})/t|^{-0.5034}.
\label{d_0_1_2}
\end{eqnarray}
The fitted curve is represented by the dotted black line in Fig.~\ref{fig_dos}(c). 
For the ideal narrow-sense type-III Dirac cone described by Eq.~(\ref{model}) with 
$\alpha/(\hbar v_{\rm F})=-1$, 
the DOS follows 
$D(\varepsilon)\propto|\varepsilon-\varepsilon_{\rm F}|^{-1/2}$ \cite{KH2025}.
The fitted exponent is therefore consistent with the prediction of the continuum model, 
whereas the prefactor is modified by lattice effects.

However, the continuum-model expression does not reproduce the DOS for  $\varepsilon>\varepsilon_{\rm F}$, 
where the numerical DOS (pink line) is significantly enhanced compared with the continuum-model prediction. 
This enhancement originates from the flat dispersion along the $k_x$ direction at $k_y=\pm\pi/a$, whose energy coincides with the Dirac-point energy of the narrow-sense type-III Dirac cone. The intersections between the two sets of flat lines, shown as two white solid lines and two white dashed lines in Fig.~\ref{fig_band_cont2}(a), may play an important role in the enhancement of the DOS. A similar DOS enhancement has been reported in anisotropic superconductors with intersecting line nodes\cite{hasegawa1996}. For $t_{2a}=0$ and $t'/t=0.5$, the DOS above the Fermi energy is more strongly enhanced in the present honeycomb lattice model  than in the continuum theory or the square lattice model, both of which describe the narrow-sense type-III Dirac cone.

For $t^{\prime}/t>0.5$, the Dirac cone becomes overtilted (type-II), resulting in a nearly constant DOS around 
$\varepsilon_{\rm F}$, as shown in Figs.~\ref{fig_dos} (b) and (c). In addition, the van Hove singularity associated with the saddle point $\mathbf{k}_{\rm S1}$ in the upper band is visible.




\begin{figure}[bt]
\begin{flushleft} 
\hspace{0.5cm}(a) \end{flushleft}\vspace{-0.2cm}
\includegraphics[width=0.5\textwidth]{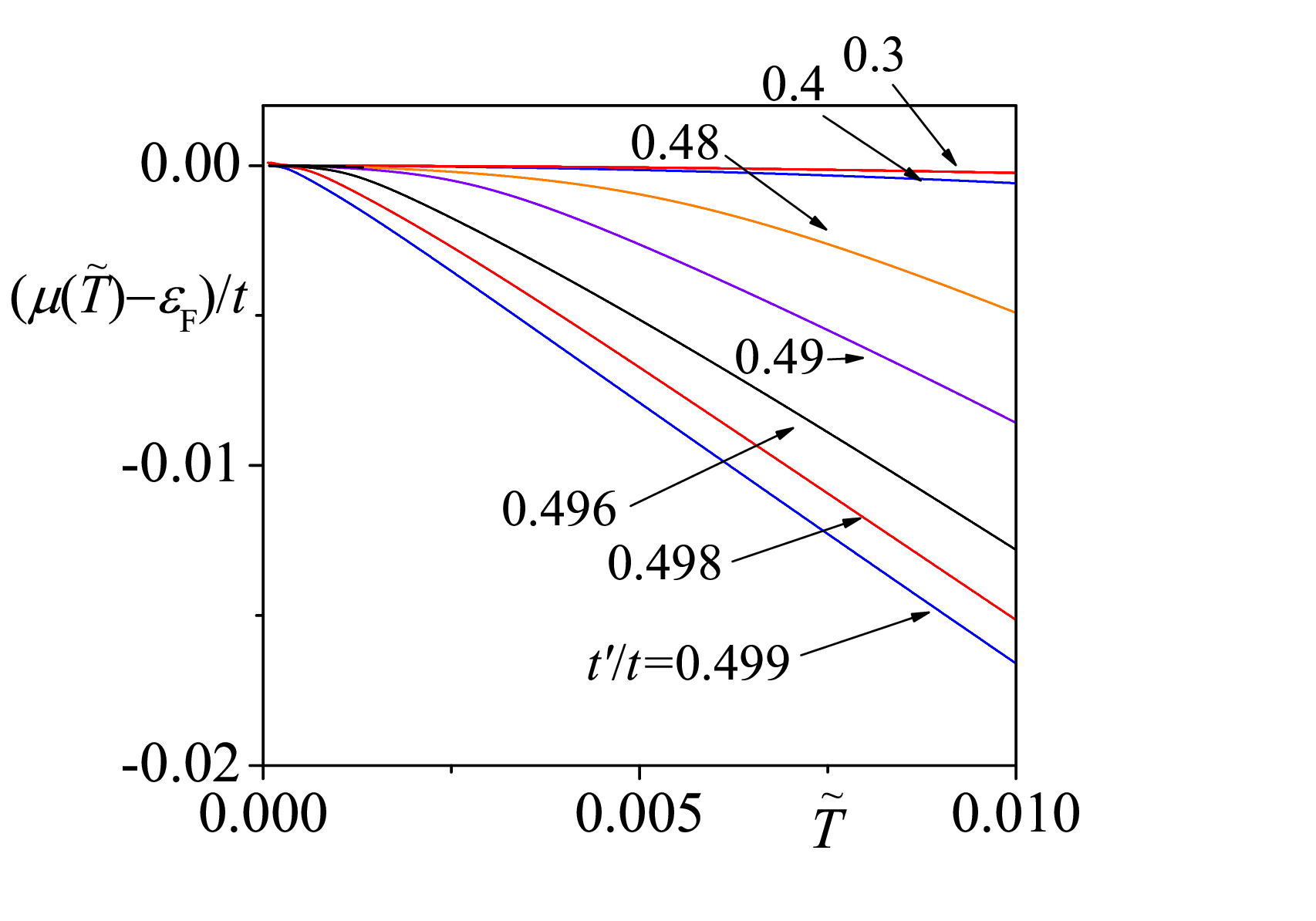}\vspace{-0.2cm}
\begin{flushleft} \hspace{0.5cm}(b) \end{flushleft}\vspace{-0.3cm}
\includegraphics[width=0.5\textwidth]{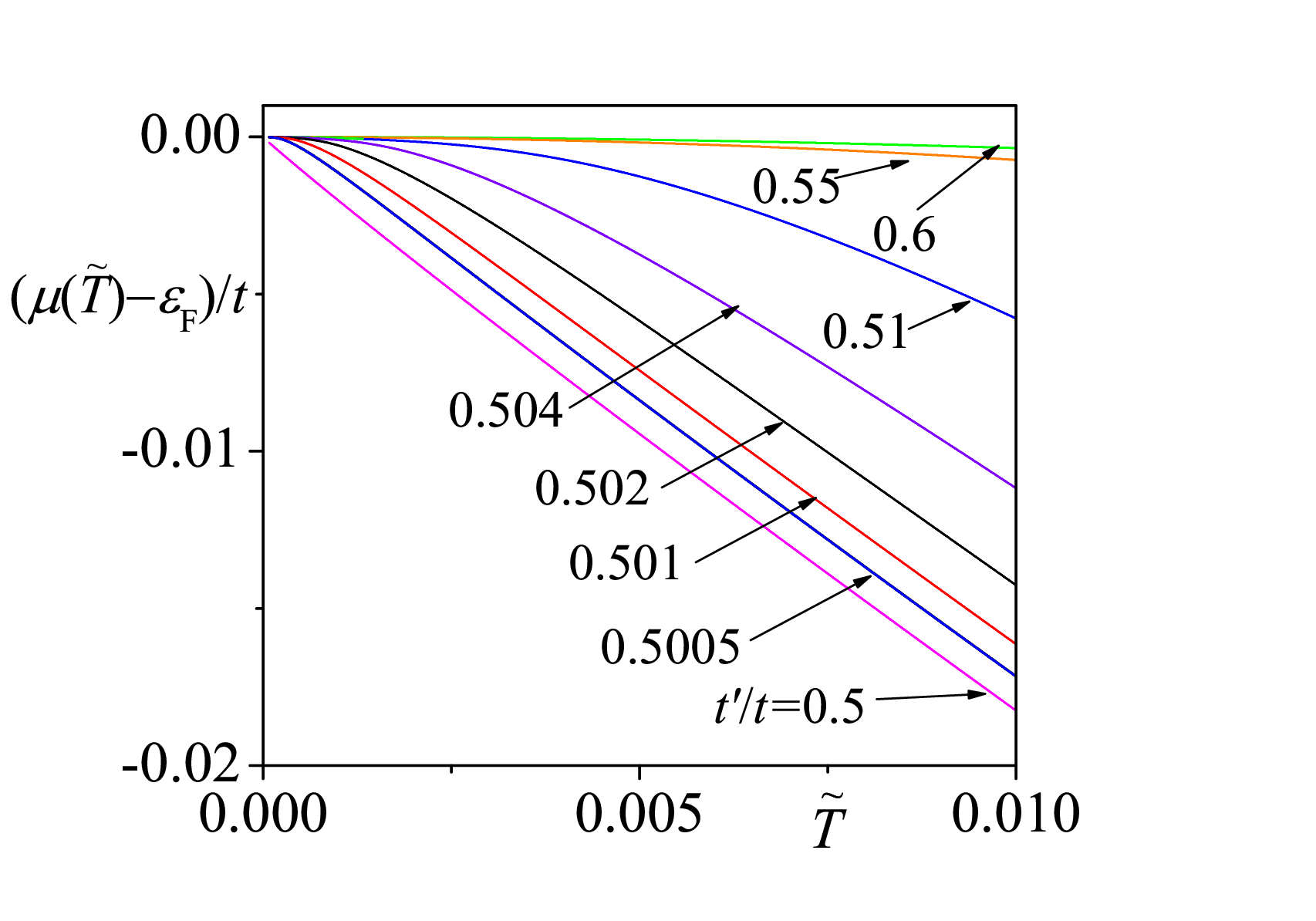}\vspace{-0.2cm}
\vspace{-0.0cm}
\caption{
 (Color online) 
The chemical potential at different $t^{\prime}/t$ values as a function of ${\tilde T}$. (a) is for type I and type III, where $t^{\prime}/t \leq 0.499$. (b) is for type II and type III, where $t^{\prime}/t \geq 0.5$.
The vertical axis shows $\mu({\tilde T})-\varepsilon_{\rm F}$, where $\varepsilon_{\rm F}$ is the zero-temperature Fermi energy. Negative values indicate that $\mu({\tilde T})$ lies below $\varepsilon_{\rm F}$, not necessarily that $\mu({\tilde T})$ itself is negative.
}
\label{fig_4}
\end{figure}

\begin{figure}[bt]
\begin{flushleft} 
\hspace{0.5cm}(a) 
\end{flushleft}\vspace{-0.2cm}
\includegraphics[width=0.51\textwidth]{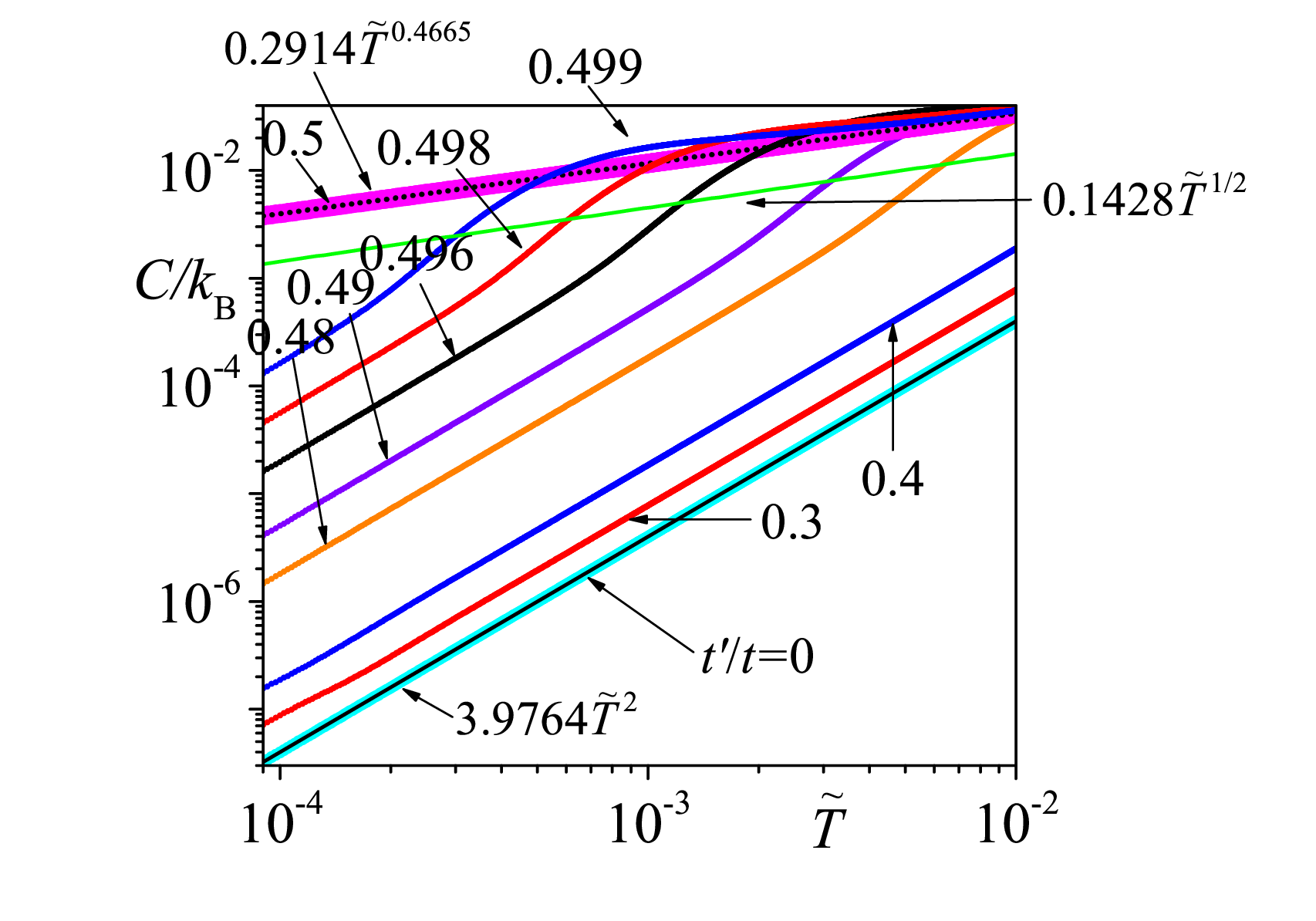}\vspace{-0.2cm}
\begin{flushleft} \hspace{0.5cm}(b) \end{flushleft}\vspace{-0.3cm}
\includegraphics[width=0.51\textwidth]{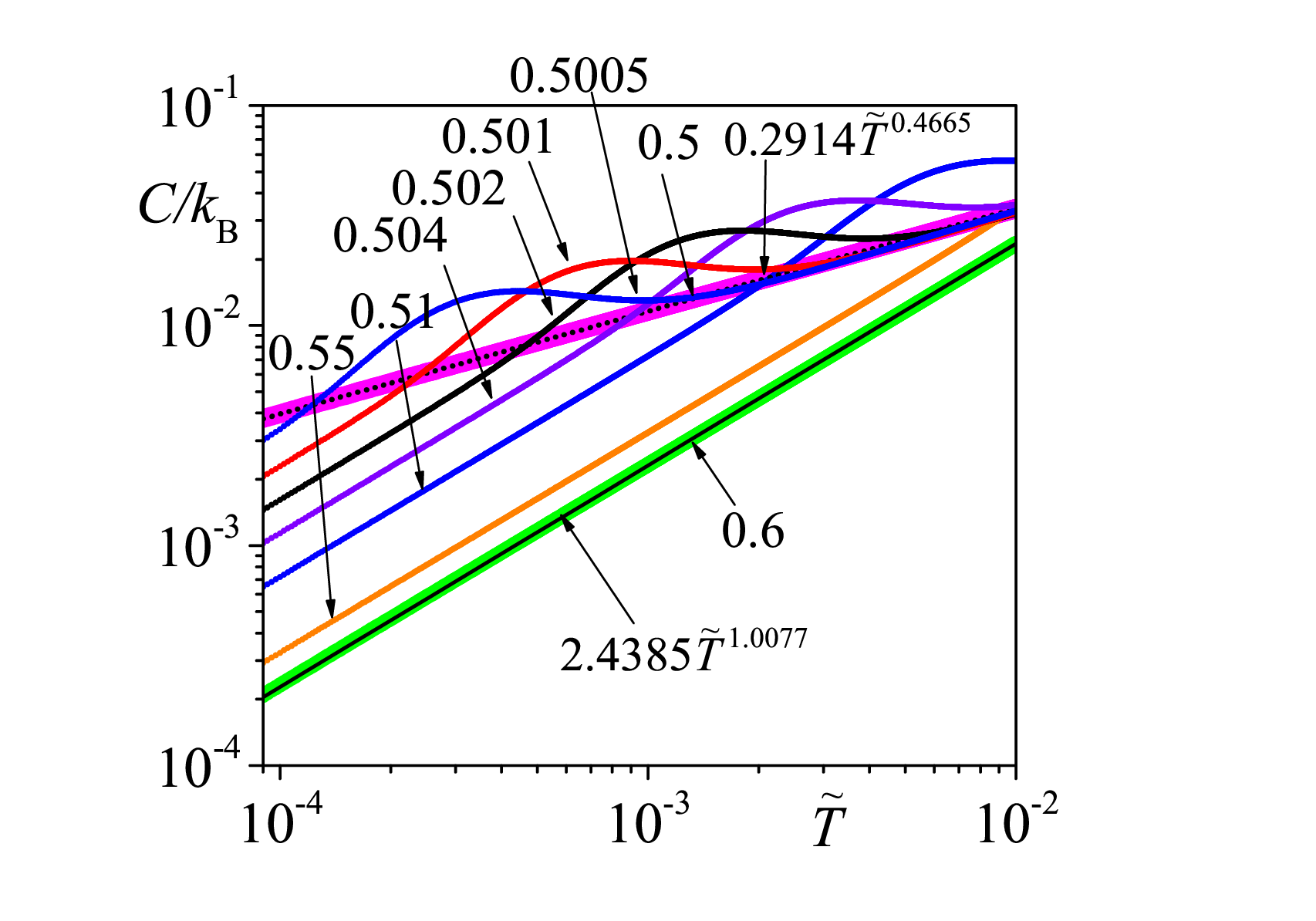}\vspace{-0.2cm}
\caption{(Color online) 
The electronic specific heats as a function of ${\tilde T}$ from numerical calculations. The values of $t^{\prime}$ in (a) and (b) are corresponding to those in Figs. \ref{fig_4} (a) and (b), respectively.  Eq. (\ref{C_D}) in (a) and  $C=2.4385{\tilde T}^{1.0077}$ in (b) are 
 represented by black lines.
}
\label{fig_5}
\end{figure}

\section{Electronic specific heat in 2D honeycomb lattice}
\label{ES_2D}

The present thermodynamic analysis is restricted to equilibrium states at positive temperatures and at half filling. The internal energy per site, $U$, at temperature $T$ is given by
\begin{equation}
U = \frac{1}{N} \sum_{i=\pm} \sum_{\mathbf{k}}
\varepsilon(i,\mathbf{k})
f[\varepsilon(i,\mathbf{k})],
\label{U_1}
\end{equation}
where $\varepsilon(i,\mathbf{k})$ denotes the energy eigenvalue of band $i$, with $+$ and $-$ corresponding to the upper and lower bands, respectively, and $f[\varepsilon(i,\mathbf{k})]$ is the Fermi-Dirac distribution function with chemical potential $\mu$. In the present study, electron-electron interactions are neglected, and the thermodynamic properties are calculated based on the single-particle band structure. The electronic specific heat at constant volume and fixed filling is obtained as
\begin{equation}
C = \frac{dU}{dT},
\label{Ce}
\end{equation}
where the chemical potential $\mu$ is determined at each temperature to keep the filling fixed. 
When the spin degree of freedom is taken into account, an additional factor of 2 should be multiplied on the right-hand side of Eq.~(\ref{Ce}).

The non-interacting approximation provides a starting point for investigating the effects of the band structure on the DOS and electronic specific heat. However, near the critical point, where the DOS is strongly enhanced near the Fermi energy, electron-electron interactions may become important and modify the low-energy electronic states and the temperature dependence of the specific heat. Investigating these effects is beyond the scope of the present study.

When particle-hole symmetry is present, the chemical potential at half
filling is independent of $T$. In this case, by using  the following generalized DOS with particle-hole symmetry:
\begin{align}
D(\varepsilon)=ND_{0}|\varepsilon-\varepsilon_{\rm F}|^{\eta}, \label{D1}
\end{align}
where $D_{0}$ and $\eta$ are constants, 
we derived the general expression\cite{KH2025} for $C$ as
\begin{align}
C = 2 k_{\rm B} D_0 (\eta + 2) t^{\eta+1}\left(1 - 2^{-(\eta + 1)}\right) \zeta(\eta + 2) \Gamma(\eta + 2)  \tilde{T}^{\eta+1}, \label{CT}
\end{align}
where $\tilde{T}=k_{\rm B}T/t$ is the dimensionless temperature.

By using the DOS of Eq. (\ref{d_0_1_1}),  the electronic specific heat is given  by
\begin{eqnarray}
C&=&\frac{6\sqrt{3}}{\pi}\zeta(3)k_{\rm B}\tilde{T}^{2} \nonumber \\
&\simeq&3.9764 k_{\rm B}\tilde{T}^{2} \ \mbox{for type-I.}\label{C_D} 
\end{eqnarray}

Based on Eq. (\ref{d_0_1_2}),  the approximate DOS for narrow-sense type-III case in the present honeycomb lattice model is given by 
\begin{eqnarray}
D(\varepsilon)
=\frac{0.0702N}{t}
|(\varepsilon-\varepsilon_{\rm F})/t|^{-1/2},
\label{d_0_1_22}
\end{eqnarray}
where the exponent is fixed to the theoretical value of $-1/2$. 
The corresponding electronic specific heat is then given by
\begin{eqnarray}
C 
&\simeq& 0.1428k_{\rm B} \tilde{T}^{1/2}. \label{C_type3}
 \end{eqnarray}

When $t^{\prime}/t\neq 0$, the particle-hole symmetry is broken in Eq.~(\ref{g1}). In this case, at half filling ($\nu=1/2$), the chemical potential $\mu({\tilde T})$ is determined from
\begin{equation}
\nu =\frac{1}{N} \sum_{i=\pm} \sum_{{\bf k}}
f[\varepsilon(i,{\bf k})].
\label{mu_N}
\end{equation}
The ${\tilde T}$-dependence of $\mu$ is shown in Fig.~\ref{fig_4}.

The ${\tilde T}$-dependence of $\mu$ arises from the particle-hole asymmetry of the DOS. In particular, near $t'/t=0.5$, the DOS on the higher-energy side of the Fermi energy is strongly enhanced by the additional flat lines in the upper band. Consequently, if the chemical potential were held fixed at $\varepsilon_{\rm F}$, the numbers of thermally excited electrons above $\varepsilon_{\rm F}$ and holes below it would be unequal. To maintain half filling, the chemical potential shifts downward as the temperature increases. The enhanced DOS and its particle-hole asymmetry near the critical point make this shift particularly significant.

The numerically calculated electronic specific heat $C$ is shown in Fig. \ref{fig_5}. 
The ${\tilde T}$-dependence of $C$ at $t^{\prime}/t=0$ [blue line in Fig. \ref{fig_5} (a)] 
agrees with Eq. (\ref{C_D}) [black line in Fig. \ref{fig_5} (a)]. At $t^{\prime}/t=0.5$ (narrow-sense type-III),  
the ${\tilde T}$-dependence of $C$, represented by the pink line, is fitted well by $C=0.2914{\tilde T}^{0.4665}$ 
in the range of 
$10^{-4}\leq {\tilde T}\leq 10^{-2}$. The magnitude of $C$ is much larger  than that predicted by Eq.~(\ref{C_type3}), shown by the green line in 
Fig.~\ref{fig_5}(a), and its exponent slightly deviates from 1/2. The enhanced magnitude of $C$ is attributed to the 
pronounced enhancement of the DOS near the Fermi energy, as shown in Fig.~\ref{fig_dos}(c). The slight deviation of the exponent from $1/2$ may arise from the $T$-dependence of the chemical potential associated with the particle-hole asymmetry of the lattice dispersion.

At $t^{\prime}/t=0.49, 0.496$, and $0.498$, the $\tilde T$-dependence of $C$ exhibits a crossover from the type-I behavior $C\propto\tilde T^{2}$ at low temperatures toward the characteristic behavior $C\propto\tilde T^{0.4665}$ found at the narrow-sense type-III point, as shown in Fig.~\ref{fig_5}(a).

At $t^{\prime}/t=0.6$, $C$ is fitted well by $C=2.4385{\tilde T}^{1.0077}$ in the range of $10^{-4}\leq\tilde T\leq10^{-2}$, as shown in Fig.~\ref{fig_5}(b). This nearly linear temperature dependence is consistent with the nearly constant DOS around the Fermi energy in the type-II regime. On the other hand, for $t^{\prime}/t=0.5005, 0.501, 0.502, 0.504$, and $0.51$, the $\tilde T$-dependence of $C$ exhibits a crossover from the type-II behavior $C\propto\tilde T$ at low temperatures toward the characteristic behavior $C\propto\tilde T^{0.4665}$ found at the narrow-sense type-III point. These results indicate that the characteristic specific-heat behavior associated with the narrow-sense type-III point extends over a finite temperature range on both sides of the critical point.

\section{Conclusions}

In this study, we demonstrated that type-I, type-II, and narrow-sense type-III Dirac cones can be realized in a two-dimensional honeycomb lattice at half filling by tuning the ratio $t^{\prime}/t$ of the anisotropic next-nearest-neighbor hoppings. For $t_{2a}=0$, $t_{2b}=t_{2c}=t'$, and $0\leq t'/t\leq 0.6$, type-I Dirac cones appear for $0 \le t^{\prime}/t < 0.5$, type-II Dirac cones for $t^{\prime}/t > 0.5$, and narrow-sense type-III Dirac cones are realized at the critical point $t^{\prime}/t = 0.5$. At this critical point, the dispersion becomes flat along the direction connecting the two Dirac points.  In artificial honeycomb lattices, previous realizations \cite{Mil} of type-III Dirac cones relied on multiorbital systems. In contrast, the present model demonstrates narrow-sense type-III Dirac cones in a single-orbital honeycomb lattice.

We also identified a distinctive feature of the present model: at the critical point $t^{\prime}/t=0.5$, additional flat lines in the upper band lie at the Fermi energy. These flat lines, which are absent in the square lattice model studied in Ref.~\cite{Ogata2025}, contribute to the pronounced enhancement of the DOS above the Fermi energy.  This enhancement may also be related to the intersections between these additional flat lines and the flat lines connecting the two Dirac points. Consequently, the electronic specific heat is significantly enhanced. In addition, the $\Gamma$ point becomes a saddle point in the upper band for $1/6 < t^{\prime}/t < 0.5$, giving rise to an additional van Hove singularity.

The present results highlight how the coexistence of narrow-sense type-III Dirac-cone dispersion and additional flat regions at the Fermi energy can enhance the DOS and electronic thermodynamic properties.  Artificial honeycomb lattices may provide a promising platform for exploring analogous band structures, since nearest- and next-nearest-neighbor couplings can be engineered through lattice design \cite{Lu}. In particular, controlled anisotropy in the next-nearest-neighbor hopping may enable the realization of narrow-sense type-III Dirac cones and additional flat lines. Although the electronic specific heat studied here is not directly applicable to photonic lattices, artificial photonic systems could provide a means of investigating analogous band structures, flat-dispersion features, and the associated enhancement of the photonic density of states.

\section*{Acknowledgement}
One of the authors (K.K.) acknowledges valuable discussions with N. Etou and M. Kawano.

\appendix
%





\input{supplement_body.tex}

\end{document}

%% file: supplement_body.tex

\section{Supplemental Material}
\subsection{Saddle-point analysis}
\label{SM_saddle}

The Hessian matrix of a function $f(k_x,k_y)$ is defined as
\begin{equation}
H=
\begin{pmatrix}
f_{xx} & f_{xy}\\
f_{yx} & f_{yy}
\end{pmatrix},
\end{equation}
where
\begin{eqnarray}
f_{xx}
=
\frac{\partial^2 f}{\partial k_x^2},
\ 
f_{xy}
=
\frac{\partial^2 f}{\partial k_x\partial k_y}, \ 
f_{yx}
=
\frac{\partial^2 f}{\partial k_y\partial k_x}, \ 
f_{yy}
=
\frac{\partial^2 f}{\partial k_y^2}.
\end{eqnarray}
When the second-order partial derivatives of $f$ are continuous, $f_{xy}=f_{yx}$. The determinant of the Hessian is then
\begin{equation}
\det H
=
f_{xx}f_{yy}-f_{xy}^2.
\end{equation}

At a stationary point ($f_x=f_y=0$), the Hessian determines the
local character of the point when $\det H\neq0$. If $\det H<0$, the point is a saddle point.
If $\det H>0$, the point is a local minimum for $f_{xx}>0$
and a local maximum for $f_{xx}<0$. When $\det H=0$, the Hessian test is inconclusive.

We analyze the saddle points of the energy dispersion
$\varepsilon(\pm,\mathbf{k})$ given in the main text. For $t_{2a}=0$ and $t_{2b}=t_{2c}=t^{\prime}$, the energy dispersion is
given by 
\begin{eqnarray}
\varepsilon(\pm,\mathbf{k})
&=&
\varepsilon_2(\mathbf{k})
\pm\varepsilon_1(\mathbf{k}),
\end{eqnarray}
where
\begin{eqnarray}
\varepsilon_1(\mathbf{k})
&=&
t\bigg[
3+4\cos\left(\frac{\sqrt{3}}{2}ak_x\right)
\cos\left(\frac{1}{2}ak_y\right)
+2\cos(ak_y)
\bigg]^{1/2},
\nonumber\\
\varepsilon_2(\mathbf{k})
&=&
-4t^{\prime}
\cos\left(\frac{\sqrt{3}}{2}ak_x\right)
\cos\left(\frac{1}{2}ak_y\right).
\label{SM_energy}
\end{eqnarray}

The energies at the points ($\mathbf{k}_{\rm S0}$, $\mathbf{k}_{\rm S1}$, $\mathbf{k}_{\rm S2}$, and $\mathbf{k}_{\rm S3}$) are
\begin{eqnarray}
\varepsilon(+, \mathbf{k}_{\rm S0})
&=&
3t-4t^{\prime},
\label{ks0}
\\
\varepsilon(\pm,\mathbf{k}_{\rm S1})
&=&
\varepsilon(\pm,\mathbf{k}_{\rm S3})
=\pm t,
\label{ks13_energy}
\\
\varepsilon(\pm,\mathbf{k}_{\rm S2})
&=&
\pm t+4t^{\prime}.
\label{ks2_energy}
\end{eqnarray}

\subsubsection{Analysis at the $\Gamma$ point}

We first consider the $\Gamma$ point,
\begin{equation}
\mathbf{k}_{\rm S0}=(0,0).
\end{equation}
Direct differentiation of $\varepsilon(\pm,\mathbf{k})$ gives
\begin{eqnarray}
\left.
\frac{\partial\varepsilon(\pm,\mathbf{k})}{\partial k_x}
\right|_{\mathbf{k}=\mathbf{k}_{\rm S0}}
&=&0,
\nonumber\\
\left.
\frac{\partial\varepsilon(\pm,\mathbf{k})}{\partial k_y}
\right|_{\mathbf{k}=\mathbf{k}_{\rm S0}}
&=&0.
\end{eqnarray}
Thus, $\mathbf{k}_{\rm S0}$ is a stationary point in both bands.

The second derivatives at $\mathbf{k}_{\rm S0}$ are
\begin{eqnarray}
f_{xx}^{(0,\pm)}
&=&
\left.
\frac{\partial^2\varepsilon(\pm,\mathbf{k})}
{\partial k_x^2}
\right|_{\mathbf{k}=\mathbf{k}_{\rm S0}}
=
\frac{a^2}{2}
\left(\mp t+6t^{\prime}\right),
\nonumber\\
f_{xy}^{(0,\pm)}
&=&
\left.
\frac{\partial^2\varepsilon(\pm,\mathbf{k})}
{\partial k_x\partial k_y}
\right|_{\mathbf{k}=\mathbf{k}_{\rm S0}}
=0,
\nonumber\\
f_{yy}^{(0,\pm)}
&=&
\left.
\frac{\partial^2\varepsilon(\pm,\mathbf{k})}
{\partial k_y^2}
\right|_{\mathbf{k}=\mathbf{k}_{\rm S0}}
=
\frac{a^2}{2}
\left(\mp t+2t^{\prime}\right).
\end{eqnarray}
Therefore,
\begin{equation}
H_{0,\pm}
=
\begin{pmatrix}
f_{xx}^{(0,\pm)} & 0\\
0 & f_{yy}^{(0,\pm)}
\end{pmatrix},
\end{equation}
and
\begin{equation}
\det H_{0,\pm}
=f_{xx}^{(0,\pm)}f_{yy}^{(0,\pm)}
\end{equation}

For $t^{\prime}/t\geq0$, both
$f_{xx}^{(0,-)}>0$ and $f_{yy}^{(0,-)}>0$.
Thus, $\mathbf{k}_{\rm S0}$ is always a local minimum
in the lower band. 

In the upper band, for $0\leq t^{\prime}/t<1/6$, both $f_{xx}^{(0,+)}< 0$ and $f_{yy}^{(0,+)}<0$,
and $\mathbf{k}_{\rm S0}$ is a local maximum. For $1/6<t^{\prime}/t<1/2$, $f_{xx}^{(0,+)}>0$ whereas $f_{yy}^{(0,+)}<0$.
Consequently, $\det H_{0,+}<0$, and $\mathbf{k}_{\rm S0}$ is a saddle point in the upper band. For $t^{\prime}/t>0.5$, $f_{xx}^{(0,+)}>0$ and $f_{yy}^{(0,+)}>0$, and $\mathbf{k}_{\rm S0}$ becomes a local minimum in the upper band.

At $t^{\prime}/t=1/6$, $f_{xx}^{(0,+)}=0$, whereas at
$t^{\prime}/t=0.5$, $f_{yy}^{(0,+)}=0$.
In both cases, $\det H_{0,+}=0$, and the Hessian test is
therefore inconclusive. At $t^{\prime}/t=1/6$, expanding
$\varepsilon(+,\mathbf{k})$ around the $\Gamma$ point up to fourth order gives 
\begin{eqnarray}
\varepsilon(+,\mathbf{k})
\simeq
\frac{7}{3}t
-\frac{a^2t}{6}k_y^2
-\frac{a^4t}{96}k_x^4
-\frac{a^4t}{48}k_x^2k_y^2+\frac{a^4t}{288}k_y^4.
\end{eqnarray}
For sufficiently small $\mathbf{k}$, the negative quadratic term in
$k_y$ dominates the positive quartic term in $k_y$, while all terms
containing $k_x$ are nonpositive. Hence, $\varepsilon(+,\mathbf{k})<7t/3$
in  any sufficiently small region around $\mathbf{k}=\mathbf{0}$.
Thus, $\mathbf{k}_{\rm S0}$ is a local maximum in the upper band.

At $t^{\prime}/t=0.5$, the dispersion is flat along the $k_y$ direction through $\mathbf{k}_{\rm S0}$, as shown in Fig.~7(a).
Thus, the energy remains constant along this direction, while
it increases for small deviations in the $k_x$ direction.
Therefore, $\mathbf{k}_{\rm S0}$ is a degenerate local minimum.

\subsubsection{Analysis at $\mathbf{k}_{\rm S1}$ and
$\mathbf{k}_{\rm S3}$}

We next consider
\begin{equation}
\mathbf{k}_{\rm S1}
=
\left(\frac{\pi}{\sqrt{3}a},\frac{\pi}{a}\right).
\end{equation}
Direct differentiation gives
\begin{eqnarray}
\left.
\frac{\partial\varepsilon(\pm,\mathbf{k})}{\partial k_x}
\right|_{\mathbf{k}=\mathbf{k}_{\rm S1}}
&=&0,
\nonumber\\
\left.
\frac{\partial\varepsilon(\pm,\mathbf{k})}{\partial k_y}
\right|_{\mathbf{k}=\mathbf{k}_{\rm S1}}
&=&0.
\end{eqnarray}
Thus, $\mathbf{k}_{\rm S1}$ is a stationary point in both bands.

The second derivatives are
\begin{eqnarray}
f_{xx}^{(1,\pm)}
&=&0,
\nonumber\\
f_{xy}^{(1,\pm)}
&=&
\pm\frac{\sqrt{3}}{2}a^2t
-\sqrt{3}a^2t^{\prime},
\nonumber\\
f_{yy}^{(1,\pm)}
&=&
\pm a^2t.
\end{eqnarray}
Thus,
\begin{equation}
H_{1,\pm}
=
\begin{pmatrix}
0&
f_{xy}^{(1,\pm)}
\\
f_{xy}^{(1,\pm)}
&
f_{yy}^{(1,\pm)}
\end{pmatrix},
\end{equation}
and
\begin{equation}
\det H_{1,\pm}
=
-{f_{xy}^{(1,\pm)}}^2
\end{equation}

For $0\leq t^{\prime}/t<0.5$,
$\det H_{1,\pm}<0$ for both bands.
Therefore, $\mathbf{k}_{\rm S1}$ is a saddle point in both bands. 

At $t^{\prime}/t=0.5$, for the upper band, 
\begin{equation}
f_{xy}^{(1,+)}=0,
\qquad
\det H_{1,+}=0.
\end{equation}
Thus, the Hessian test is inconclusive for the upper band. 
As shown in Fig.~7(a), the dispersion is flat along the $k_x$ direction through $\mathbf{k}_{\rm S1}$. 
Along this direction, the energy remains constant, while
it increases for small deviations in the $k_y$ direction.
Thus, $\mathbf{k}_{\rm S1}$ is a degenerate local minimum in the upper band. 

For the lower band,
\begin{equation}
f_{xy}^{(1,-)}=-\sqrt{3}a^2t,
\end{equation}
and hence
\begin{equation}
\det H_{1,-}=-3a^4t^2<0.
\end{equation}
Therefore, $\mathbf{k}_{\rm S1}$ remains a saddle point in the lower band.

For $t^{\prime}/t>0.5$,
$\det H_{1,\pm}<0$ again, and $\mathbf{k}_{\rm S1}$
is a saddle point in both bands.

Similarly, we consider
\begin{equation}
\mathbf{k}_{\rm S3}
=
\left(\frac{\pi}{\sqrt{3}a},-\frac{\pi}{a}\right).
\end{equation}
The first derivatives vanish:
\begin{eqnarray}
\left.
\frac{\partial\varepsilon(\pm,\mathbf{k})}{\partial k_x}
\right|_{\mathbf{k}=\mathbf{k}_{\rm S3}}
&=&0,
\nonumber\\
\left.
\frac{\partial\varepsilon(\pm,\mathbf{k})}{\partial k_y}
\right|_{\mathbf{k}=\mathbf{k}_{\rm S3}}
&=&0.
\end{eqnarray}
The second derivatives are
\begin{eqnarray}
f_{xx}^{(3,\pm)}
&=&0,
\nonumber\\
f_{xy}^{(3,\pm)}
&=&
\mp\frac{\sqrt{3}}{2}a^2t
+\sqrt{3}a^2t^{\prime}=-f_{xy}^{(1,\pm)},
\nonumber\\
f_{yy}^{(3,\pm)}
&=&
\pm a^2t.
\end{eqnarray}
Thus,
\begin{equation}
H_{3,\pm}
=
\begin{pmatrix}
0&
f_{xy}^{(3,\pm)}
\\
f_{xy}^{(3,\pm)}
&
f_{yy}^{(3,\pm)}
\end{pmatrix},
\end{equation}
and we obtain 
\begin{equation}
\det H_{3,\pm}
=
\det H_{1,\pm}.
\end{equation}
Thus, $\mathbf{k}_{\rm S3}$ is a saddle point in the same bands
as $\mathbf{k}_{\rm S1}$. Furthermore, the two points are degenerate in energy:
\begin{equation}
\varepsilon(\pm,\mathbf{k}_{\rm S1})
=
\varepsilon(\pm,\mathbf{k}_{\rm S3})
=
\pm t.
\label{ks13}
\end{equation}

\subsubsection{Analysis at $\mathbf{k}_{\rm S2}$}

We finally consider
\begin{equation}
\mathbf{k}_{\rm S2}
=
\left(\frac{2\pi}{\sqrt{3}a},0\right).
\end{equation}
The first derivatives vanish:
\begin{eqnarray}
\left.
\frac{\partial\varepsilon(\pm,\mathbf{k})}{\partial k_x}
\right|_{\mathbf{k}=\mathbf{k}_{\rm S2}}
&=&0,
\nonumber\\
\left.
\frac{\partial\varepsilon(\pm,\mathbf{k})}{\partial k_y}
\right|_{\mathbf{k}=\mathbf{k}_{\rm S2}}
&=&0.
\end{eqnarray}
Thus, $\mathbf{k}_{\rm S2}$ is a stationary point in both bands.
Direct differentiation gives
\begin{eqnarray}
f_{xx}^{(2,\pm)}
&=&
\pm\frac{3a^2}{2}t
-3a^2t^{\prime},
\nonumber\\
f_{xy}^{(2,\pm)}
&=&0,
\nonumber\\
f_{yy}^{(2,\pm)}
&=&
\mp\frac{a^2}{2}t
-a^2t^{\prime}.
\end{eqnarray}
Therefore,
\begin{equation}
H_{2,\pm}
=
\begin{pmatrix}
\displaystyle
f_{xx}^{(2,\pm)}
&0\\[2mm]
0&
\displaystyle
f_{yy}^{(2,\pm)}
\end{pmatrix},
\end{equation}
and
\begin{equation}
\det H_{2,\pm}
=
f_{xx}^{(2,\pm)}f_{yy}^{(2,\pm)}.
\end{equation}

For
$0\leq t^{\prime}/t<0.5$,
the diagonal elements have opposite signs in both bands.
Therefore,
\begin{equation}
\det H_{2,\pm}<0,
\end{equation}
and $\mathbf{k}_{\rm S2}$ is a saddle point in both bands.

At $t^{\prime}/t=0.5$,
\begin{equation}
f_{xx}^{(2,+)}=0,
\qquad
f_{yy}^{(2,-)}=0,
\end{equation}
and hence
\begin{equation}
\det H_{2,+}
=
\det H_{2,-}
=0.
\end{equation}
The Hessian test is therefore inconclusive for both bands. For the upper band, expanding $\varepsilon(+,\mathbf{k})$ around $\mathbf{k}_{\rm S2}$ gives  
\begin{eqnarray}
\varepsilon(+,\mathbf{k})&\simeq&3t
-\frac{a^2t}{2}k_y^2
-\frac{9a^4t}{32}(k_x-\frac{2\pi}{\sqrt{3}a})^4\nonumber \\
&-&\frac{3a^4t}{16}(k_x-\frac{2\pi}{\sqrt{3}a})^2k_y^2
-\frac{3a^4t}{64}k_y^4.
\end{eqnarray}
Thus,
$\varepsilon(+,2\pi/(\sqrt{3}a),k_y)<3t$
and
$\varepsilon(+,k_x,0)<3t$
for sufficiently small deviations from $\mathbf{k}_{\rm S2}$.
Therefore, at $t^{\prime}/t=0.5$,
$\mathbf{k}_{\rm S2}$ is a local maximum in the upper band. 
For the lower band, 
the dispersion is flat along the $k_y$ direction through $\mathbf{k}_{\rm S2}$, as shown in Fig.~7(b). Along this direction, the energy remains constant, while it decreases for small deviations in the $k_x$ direction. Thus, at $t^{\prime}/t=0.5$, $\mathbf{k}_{\rm S2}$ is a degenerate local maximum in the lower band.

For $t^{\prime}/t>0.5$, both diagonal elements are negative in the upper and lower bands. Therefore,
\begin{equation}
\det H_{2,\pm}>0,
\end{equation}
and $\mathbf{k}_{\rm S2}$ is a local maximum in both bands.



\begin{figure}[bt]
\includegraphics[width=0.5\textwidth]{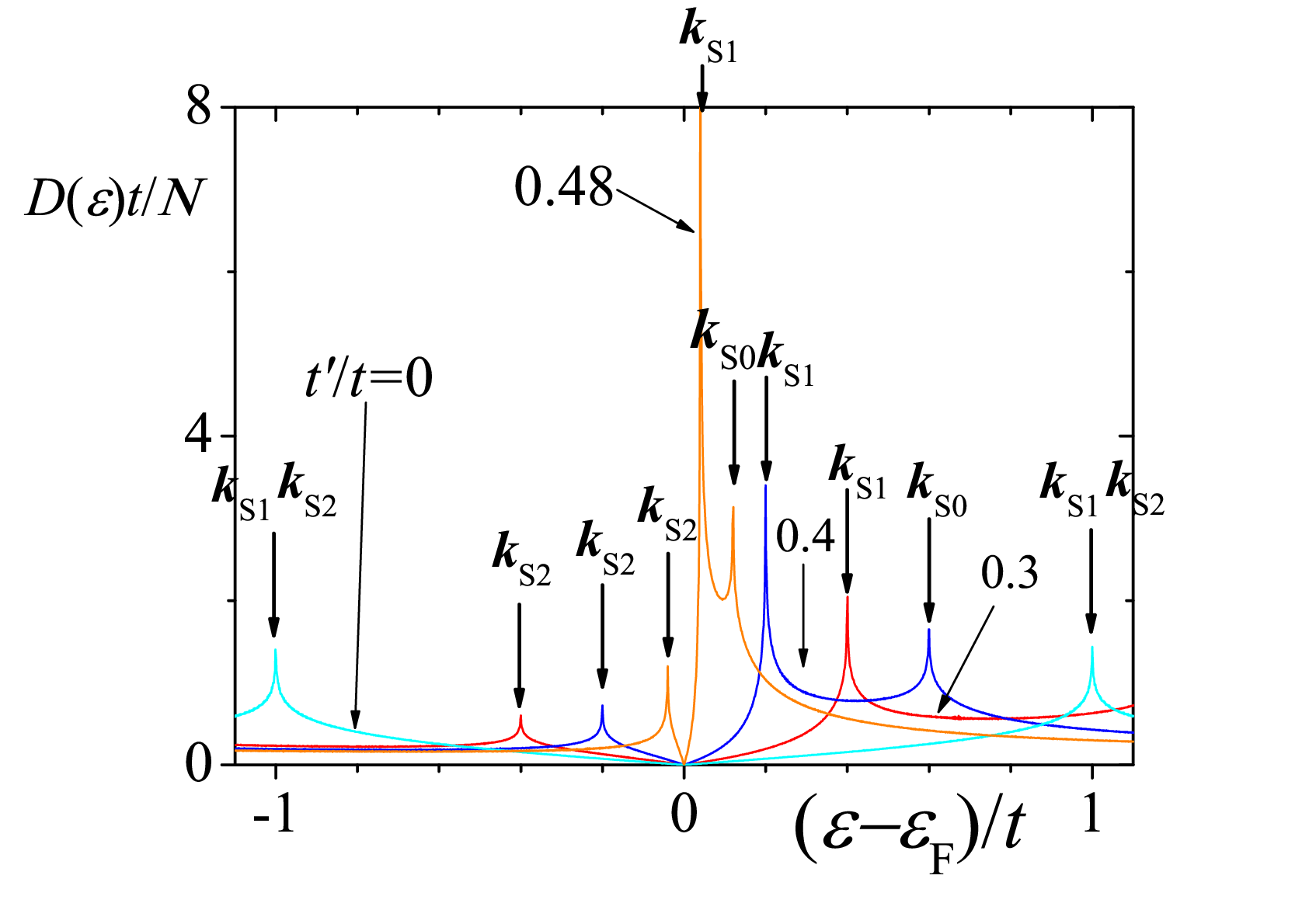}\vspace{-0.2cm}
\caption{(Color online) 
The numerically obtained DOS for several values of $t^{\prime}/t$: $0$, $0.3$, $0.4$, and $0.48$.  
}\label{figS1}
\end{figure}

\subsection{Density of states and van Hove singularities}

The van Hove singularities associated with the saddle points are clearly visible as peaks in 
the density of states (DOS), as indicated by the arrows in Fig.~\ref{figS1}. For example, for $t^{\prime}/t=0$, peaks appear at 
$(\varepsilon(\pm, \mathbf{k}_{\rm S1})-\varepsilon_{\rm F})/t =(\varepsilon(\pm, \mathbf{k}_{\rm S2})-\varepsilon_{\rm F})/t=\pm 1$, whereas for $t^{\prime}/t=0.48$, peaks appear at 
$(\varepsilon(+, \mathbf{k}_{\rm S0})-\varepsilon_{\rm F})/t =0.12$, 
$(\varepsilon(+, \mathbf{k}_{\rm S1})-\varepsilon_{\rm F})/t =0.04$, and 
$(\varepsilon(-, \mathbf{k}_{\rm S2})-\varepsilon_{\rm F})/t =-0.04$.